\documentclass[manuscript,screen,nonacm]{acmart}

\AtBeginDocument{%
  \providecommand\BibTeX{{%
    \normalfont B\kern-0.5em{\scshape i\kern-0.25em b}\kern-0.8em\TeX}}}

\usepackage[utf8]{inputenc}
\usepackage{balance}
\usepackage{amsmath}
\usepackage{amsthm}
\usepackage{amsfonts}       
\usepackage{algorithm,algorithmic}
\usepackage{listings}
\usepackage{accents}
\usepackage{amsfonts}
\usepackage{bm}
\usepackage{multirow}

\usepackage{graphicx}
\usepackage{amsmath}
\usepackage{listings}

\usepackage[capitalise]{cleveref}

\DeclareTextFontCommand{\emph}{\em}

\newcommand{\commentout}[1]{}

\newcommand{\junk}[1]{}
\newcommand{\reals}{\mathbb{R}}

\newcommand{\RNN}{\mathit{RNN}}
\newcommand{\MLP}{\mathit{MLP}}

\theoremstyle{definition}

\newcommand{\calX}{\mathcal{X}}

\newcommand{\calQ}{\mathcal{Q}}
\newcommand{\calU}{\mathcal{U}}

\definecolor{codegreen}{rgb}{0,0.6,0}
\definecolor{codegray}{rgb}{0.5,0.5,0.5}
\definecolor{codepurple}{rgb}{0.58,0,0.82}
\definecolor{backcolour}{rgb}{0.95,0.95,0.92}

\lstdefinestyle{mystyle}{
    backgroundcolor=\color{backcolour},   
    commentstyle=\color{codegreen},
    keywordstyle=\color{magenta},
    numberstyle=\tiny\color{codegray},
    stringstyle=\color{codepurple},
    basicstyle=\ttfamily\footnotesize,
    breakatwhitespace=false,         
    breaklines=true,                 
    captionpos=b,                    
    keepspaces=true,                 
    numbers=left,                    
    numbersep=5pt,                  
    showspaces=false,                
    showstringspaces=false,
    showtabs=false,                  
    tabsize=2
}

\newtheoremstyle{TheoremNum}%
    {\topsep}{\topsep}
    {\itshape}
    {}
    {\bfseries}
    {.}
    { }
    {\thmname{#1}\thmnote{ \bfseries #3}}
\theoremstyle{TheoremNum}

\begin{document}
\title[User Simulation to Evaluate Preference Elicitation Policies]{Minimizing Live Experiments in Recommender Systems:\\
User Simulation to Evaluate Preference Elicitation Policies}

\author{Chih-Wei Hsu}
\affiliation{%
  \institution{Google Research}
  \streetaddress{1600 Amphitheatre Parkway}
  \city{Mountain View}
  \country{USA}
  \postcode{94043}
}
\email{cwhsu@google.com}

\author{Martin Mladenov}
\affiliation{%
  \institution{Google Research}
  \streetaddress{1600 Amphitheatre Parkway}
  \city{Mountain View}
  \country{USA}
  \postcode{94043}
}
\email{mmladenov@google.com}

\author{Ofer Meshi}
\affiliation{%
  \institution{Google Research}
  \streetaddress{1600 Amphitheatre Parkway}
  \city{Mountain View}
  \country{USA}
 \postcode{94043}
}
\email{meshi@google.com}

\author{James Pine}
\affiliation{%
  \institution{Google Research}
  \streetaddress{1600 Amphitheatre Parkway}
  \city{Mountain View}
  \country{USA}
  \postcode{94043}
}
\email{rubicon@google.com}

\author{Hubert Pham}
\affiliation{%
  \institution{Google Research}
  \streetaddress{1600 Amphitheatre Parkway}
  \city{Mountain View}
  \country{USA}
  \postcode{94043}
}
\email{huberpham@google.com}

\author{Shane Li}
\affiliation{%
  \institution{Google}
  \city{Mountain View}
  \country{USA}
}
\email{lishane@google.com}

\author{Xujian Liang}
\affiliation{%
  \institution{Google}
  \city{Mountain View}
  \country{USA}
}
\email{xujianliang@google.com}

\author{Anton Polishko}
\affiliation{%
  \institution{Google}
  \city{Mountain View}
  \country{USA}
}
\email{antp@google.com}

\author{Li Yang}
\affiliation{%
  \institution{Google Research}
  \streetaddress{1600 Amphitheatre Parkway}
  \city{Mountain View}
  \country{USA}
  \postcode{94043}
}
\email{lyliyang@google.com}

\author{Ben Scheetz}
\affiliation{%
  \institution{YouTube}
  \city{New York}
  \country{USA}
}
\email{bscheetz@google.com}

\author{Craig Boutilier}
\affiliation{%
  \institution{Google Research}
  \streetaddress{1600 Amphitheatre Parkway}
  \city{Mountain View}
  \country{USA}
}
\email{cboutilier@google.com}

\renewcommand{\shortauthors}{Chih-Wei Hsu et al.}

\begin{abstract}
Evaluation of policies in recommender systems typically involves A/B testing using live experiments on real users to assess a new policy's impact on relevant metrics. This ``gold standard'' comes at a high cost, however, in terms of cycle time, user cost, and potential user retention. In developing policies for  \emph{onboarding} new users, these costs can be especially problematic, since on-boarding occurs only once. In this work, we describe a simulation methodology used to augment (and reduce) the use of live experiments. We illustrate its deployment for the evaluation of \emph{preference elicitation} algorithms used to onboard new users of the YouTube Music platform. By developing counterfactually robust user behavior models, and a simulation service that couples such models with production infrastructure, we are able to test new algorithms in a way that reliably predicts their performance on key metrics when deployed live.
We describe our domain, our simulation models and platform, results of experiments and deployment, and suggest future steps needed to further realistic simulation as a powerful complement to live experiments.
%
\end{abstract}


\maketitle

\section{Introduction}
\label{sec:intro}
\emph{Recommender systems} (RSs) play a crucial role in making 
digital content accessible to users in domains such as e-commerce, news, video, among others \citep{Abel2011,Hallinan2016,Linden2003,Pal2020,Covington2016}.
RSs typically leverage past user interactions to learn about a user's preferences to improve their recommendations.
However, when information about the user's preferences is minimal, or even non-existent in the case of new users, preference prediction is especially challenging---this is often dubbed the \emph{cold-start} problem \citep{Lam2008,bobadilla2012}.
Cold-start can be addressed in various ways \cite{cf_sparsity}, but for new users
it is often natural to ask some preliminary questions about their preferences during an \emph{onboarding process}, using some form of explicit \emph{preference elicitation (PE)} \cite{keeney1993decisions,salo2001preference,AlMamunur2008}.
In this work, we address the problem of onboarding new users of the YouTube Music platform using various PE methods that adaptively, in a personalized fashion, ask a new user about their preferences for specific musical artists (see \cref{fig:TB_screen} and Sec.~\ref{sec:PE} for further details).

Developing new algorithms for any user interaction in an RS is challenging. Offline data generated by past interactions with users is commonly used to build predictive models of user responses to an RS's actions or recommendations (e.g., click-through, purchase, or engagement models)
\cite{oord_etal:nips13,covington_etal:recsys16}, and use these to devise new policies. However, because new policies often influence usage, user state, and other factors---effectively, they change the \emph{data distribution} relative to that on which models were originally trained---rarely is a new policy deployed without first running controlled A/B tests in \emph{live experiments (LEs)} with real users to assess its impact on key long-term RS metrics \cite{kohavi_controlledAB:DMKD2009,shani:RSHandbook2011}. Indeed, well-designed A/B tests are considered to be the gold standard for such evaluation.\footnote{User studies can complement such tests \cite{radensky_etal:facct23,balog_etal:sigir20,bansal_etal:hcomp19}, but we do not address this here.}

Of course, the use of LEs for A/B testing has well-known costs \cite{shani:RSHandbook2011}. Online evaluation can impose reputational and retention costs by degrading the experience of users exposed to poorly performing (experimental) policies. Statistically significant results---which place strict demands on required sample sizes
---may require non-trivial amounts of time to emerge, which can slow the algorithm development cycle.
Finally, opportunity cost is especially acute when testing onboarding methods: since users onboard only once (in contrast to, say, repeated interaction with content recommendations), a poor user experience induced by an experimental algorithm leaves no opportunity for recovery.

To alleviate these concerns, \emph{simulation} has been used increasingly for RS research and algorithm development \cite{wang_etal:recsys23, recsimNG:arxiv21,ahlgren-facebookSim:ease21,yao_halpern_etal:fates20,recsim:arxiv19}. Simulated \emph{user models} allow developers to ``observe'' how (stochastic) user responses unfold in reaction to sequences of RS actions or recommendations, nominally circumventing the out-of-distribution issues of static data sets, allowing effective assessment of new algorithms, and supporting quicker iteration without imposing the costs of LEs on users. That said, simulation is largely used for algorithm development rather than to replace or reduce LEs since simulation models tend to be relatively stylized, are often designed to predict short-term effects rather than the long-term metrics typically assessed by LEs, and are rarely connected to production infrastructure (which links other RS components, beyond the algorithm in question, that influence user behavior).



In this work, we propose a practical methodology for the use of simulation not only to help develop new RS algorithms, but to move a step closer to using simulation to reduce (and even partially replace) LEs for evaluation. We illustrate this methodology using our experience with the development, evaluation and deployment of novel onboarding algorithms for new users of YouTube Music. Along the way we outline: (a) the development of user models---in our case recurrent neural networks \cite{lstm} and transformers \cite{vaswani2017attention}---that capture the distribution of (relevant properties of) real users expected to be seen under product usage; (b) a process that ensures these models are reasonably counterfactually robust so performance predictions for \emph{new} policies can be used to guide deployment decisions \cite{recsim:arxiv19}; and (c) the framework used to run our simulated users against production infrastructure \cite{ahlgren-facebookSim:ease21} to ensure simulated metrics reflect \emph{all} product intricacies. We conclude with a discussion of some open challenges that need to be addressed to make our methodology universally applicable.


\section{Onboarding in a Music RS}
\label{sec:PE}

We begin by outlining the general onboarding interface and experience for new users of YouTube Music, and the types of algorithms---namely, \emph{preference elicitation (PE)} methods---we have developed to drive actual user interaction. The effective development and evaluation of these PE methods motivates our simulation framework and methodology.
We refer to \citet{meshi2023preference} for a more detailed description of the PE methods themselves.

We assume an RS designed to recommend items from a large content corpus. We use music recommendation as our focus, where items are music tracks. Behavior-based \emph{collaborative filtering (CF)} is a dominant approach to RS design in content domains, where methods such as matrix factorization \cite{salakhutdinov-mnih:nips07} or neural CF \cite{beutel_etal:wsdm18,yangEtAl:www20} are used to generate user and item embeddings in some latent space $\calX\subseteq\reals^d$. A user's general affinity for an item is then given by the dot product or cosine similarity between their embeddings.
We assume access to stable \emph{item embeddings} $\phi(i)\in\calX$.

CF methods require sufficient behavioral data (e.g., listens, item ratings) to generate a usable embedding
for a user; hence, 
valuable (non-generic) recommendations are not viable for new users with no interaction history.
%
One way to address this \emph{cold start problem} \cite{Lam2008,bobadilla2012} is to use PE \cite{keeney1993decisions,salo2001preference,chajewska2000making,preference:aaai02}.
For new users, PE may be used as part of an \emph{onboarding process} designed to make the service quickly usable and useful. PE methods require some class of \emph{queries} designed to elicit information about a user's preferences, a specific semantics for user responses, and a procedure for selecting queries (which often adapts to a user's previous responses). For example, PE  might ask users for individual item preferences (e.g., ``do you like track X?''), to compare two items (e.g., ``which track do you prefer, X or Y?''), or for attribute preferences (e.g., ``do you like (tracks by) artist A?'' or ``do you prefer genre G to genre H?'').

Item-based elicitation is often effective with small item corpora, or when making fine-grained distinctions when the RS already has a good understanding of a user's preferences. However, it is unsuitable for onboarding with vast corpora like those in music recommendation: the granularity of item-based PE allows precision, but renders sufficient coverage of the music domain infeasible during a time-constrained onboarding process. Thus, we focus on attribute-based elicitation, specifically, asking a user for their preferences w.r.t.\ one or more \emph{music artists}.
We assume that attributes (artists) are embedded in the same latent space as tracks so a user's artist preferences informs the RS's assessment of their track preferences.
The embedding of query (i.e., artist) $q$ is denoted by $\phi(q)\in\calX$.

\begin{figure}[t]
\begin{center}
\centerline{\includegraphics[width=0.35\columnwidth]{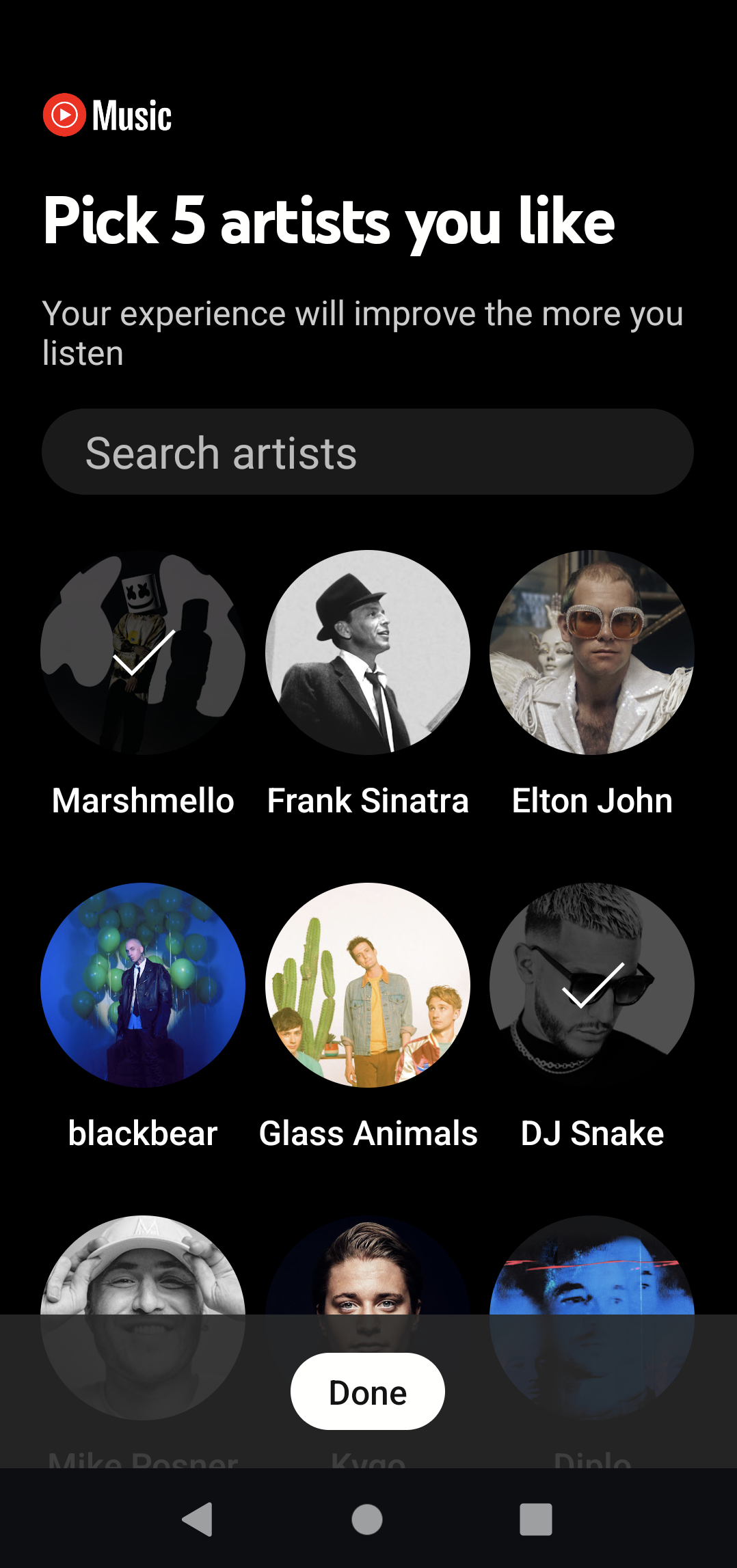}}
\caption{The artist selection interface. Users select artists they like, and skip those they don't. The scrollable interface allows for selection of as many artists as desired, with ``Done'' confirming the end of the onboarding session. The artists displayed as the user scrolls are selected dynamically given earlier selections and non-selections (or \emph{skips}).}
\label{fig:TB_screen}
\end{center}
\vskip -0.2in
\end{figure}

The interface used to elicit artist preferences is shown in Figure~\ref{fig:TB_screen}. The user can scroll to select as many artists as desired and choose ``Done'' to terminate the onboarding process.  It is important that onboarding quickly and effectively discover user preferences that support downstream recommendations, since it delays actual music consumption, and a user can terminate the process at any point.
As such, the PE strategy plays a critical role in the user experience: it impacts \emph{how much} and \emph{what type} of preference information the RS obtains, hence downstream recommendation quality and even overall RS usage. For example, if the PE algorithm shows many similar artists to the user, or artists they are not familiar with, the user might decide that onboarding is not worth their time, and abandon before providing much (or any) preference information. A strategy that does not cover the diversity of a user's musical tastes may lead to less-than-ideal initial RS performance, as may a method that does not delve deeply enough into parts of the music space important to the user.
Finally, patience or inclination toward various modes of interaction varies across users.

These considerations mean that developing a good PE algorithm for onboarding is highly non-trivial, involving tradeoffs between familiarity and novelty, coverage and ``deep dives,'' and much more. We do not provide detailed algorithm descriptions here other than to point out that our PE methods are both personalized to user context (see below) and dynamic (i.e., artists are chosen adaptively based on a user's previous selections/skips during the onboarding session). For our purposes, it is sufficient to treat various algorithms and their parameterizations as black boxes to be evaluated or compared.\footnote{The actual PE methods deployed here involve different methods of personalization , the use of information gain/value to select artists, and tradeoffs between artist coverage/diversity and depth of preference understanding. We refer to \cite{meshi2023preference} for details. For further discussion of onboarding techniques for RSs, see \cite{mcnee_onboarding:um03}, and PE, see \cite{chajewska2000making,preference:aaai02,viappiani:nips2010}.}

\section{Evaluating Preference Elicitation Policies with Simulation}
\label{sec:pe_sim}

Our goal is to effectively evaluate new PE policies for onboarding. While A/B testing of a new policy versus an incumbent using live experiments (LEs) is the usual approach, it comes with various costs, including the potential imposition on users, and the elongation of the development/testing/refinement cycle for new algorithms. Both of these costs are magnified in our onboarding setting: the first due to the one-off nature of onboarding; and the second due to the novelty of the approaches being tested relative to incumbent policies.
We consider the use of \emph{simulation} to mitigate both of these issues.
Simulation has been proposed in recent years as an important tool for RSs. It has primarily been used to reduce the second cost above, namely, algorithm-development cycle time \cite{rohde2018recogym,recsim:arxiv19,yao_halpern_etal:fates20,wang_etal:recsys23}, or analyze user dynamics \cite{ahlgren-facebookSim:ease21}. These can be especially important when testing sequential interaction involving (say) reinforcement learning \cite{chen_etal:2018top,zhao_slateRL:recsys18,slateQ:ijcai19}, or multiagent interactions \cite{mladenov_etal:icml20}.

Effective simulation for algorithm development requires one to develop synthetic \emph{user models} that capture both the distribution of relevant user characteristics, and user behavioral responses to recommender actions (e.g., recommendations, questions). These characteristic and response models need not reflect
every aspect of users; 
but when tested in a population of simulated users, they should be faithful enough to assess whether one algorithm is expected to perform better than another when deployed.
Many of the methods cited above adopt this perspective. In our case, user models reflect (conditional) user responses to questions about artist preferences, their chance of exiting the onboarding process at any stage, and the distribution of such behaviors in the new-user population.

Developing such user models is itself quite challenging. Typically, an existing RS policy has already been deployed whose data can be used to train predictive user models. Unfortunately, this static offline approach is often unable to predict behaviors when users are presented with \emph{new policies}, since new policies generate data and behaviors out-of-distribution w.r.t.\ the production policy. Without effective techniques for \emph{off-policy evaluation} \cite{doubly_robust}, these models are unlikely to be \emph{counterfactually robust}. We approach this issue by building models using data collected from several policies that we believe capture a suitable range of user behaviors. We discuss this issue further in Sec.~\ref{sec:modeling} where we describe our user models, Sec.~\ref{sec:experiments} where we validate our approach empirically, and Sec.~\ref{sec:production} where we outline additional  challenges for future research.

Apart from reducing cycle time, we also want to use simulation to predict how \emph{actual deployment} of a new algorithm in production will impact key performance metrics, including user satisfaction/value. This raises another complication: typical production RSs consist of many interacting components, all of which influence what a user sees/experiences, of which the new algorithm is just one part. A true evaluation of an algorithm's impact requires integrating it into the production stack---something that happens naturally in both LEs and deployment---and running the \emph{production RS} against simulated users. RS simulation work rarely considers the integration of synthetic or simulated users into production infrastructure---one exception is Meta's WW model \cite{ahlgren-facebookSim:ease21}, used to simulate, for example, scam and spam attacks on users. This integration is critical if simulation is to replace, or at least reduce the use of, LEs. We describe our framework for production-level simulation in Sec~\ref{sec:sim}.

\section{User Models}
\label{sec:modeling}

We now describe our user models, designed to interact with the onboarding PE module of the RS.
Our models have two requirements: generating a \emph{synthetic user population} whose distribution of relevant characteristics
reflects those of the user base; and providing a \emph{user response model} that accurately reflects user behavior with our PE methods conditioned on these characteristics.


We make two key assumptions that support our modeling approach. First, some version of the \emph{onboarding process (OP)} (Sec.~\ref{sec:PE}) has been deployed prior to the development or testing of the new methods that are the target of simulation. Often simple methods are deployed prior to developing more sophisticated algorithms. Second, most users who engaged with this original OP moved on to use the RS---in this case YouTube Music---for an extended period. Together, this provides access to a set of users $\calU$, each of whom has engaged with the existing (or prior) OP to generate \emph{OP session data}, and, following onboarding, used the RS to engage with content giving us \emph{post-onboarding usage}. Specifically, each $u\in\calU$ is associated with a single OP session $S^u$ and static user state (or \emph{context}) $C^u$.  We elaborate on the form of this data below.

The new OP that we wish to test is driven by one of several PE algorithms being considered for evaluation. These algorithms adapt their artist selection to the previous selections/skips of the user.\footnote{Our PE methods are also personalized given prior music engagement by the user, but we set this aside in our treatment here.} When presented with an artist, a user makes two distinct binary choices: (i) select or skip that artist; and (ii) terminate or continue with the OP.  Simulating a user engaging with the OP requires generating these two 
actions at each step.
Action sampling is conditioned on (a) the \emph{static (or prior) user state}; and (b) the user's OP \emph{session history} so far. The interaction of a synthetic user $u$ with the OP proceeds through a sequence of $t > 0$ turns as follows:
\begin{enumerate}
    \item RS (PE module) queries $u$ with an artist.
    \item $u$ responds (selects/skips, terminates/continues).
    \item $u$'s context (latent state) is updated.
    \item RS (PE module) updates its beliefs about the user given the response (note: we treat this as a black box). 
\end{enumerate}
This repeats until $u$ terminates the session. 

\subsection{User State Generator}
\label{sec:session_gen}

The first component of our generative user model comprises a distribution $P(C^u)$ of \emph{static user states}, those user characteristics that drive a user's behavior when engaging with the OP. Our \emph{user state generator} divides this user state into \emph{observable context} $C^u_o$ and \emph{latent state} $C^u_\ell$.
Observable context consists of factors that are directly observable and (potentially) influence a user's musical tastes and OP interactions---these might include geography, device type (e.g., mobile vs.\ desktop, OS type), and other such features.\footnote{Geography is strongly correlated with musical preferences, while device type influences the UI layout, hence OP interactions.} The distribution of these features is readily available in data (see below), and the OP's PE algorithm has access to these features.

Perhaps more critical is the user's latent state, the most important element of which is the user's \emph{true} underlying musical preferences. These are, of course, not observable; moreover, for new users, there is no data to support its estimation. However, by assumption, we have a large number of users who have not only used the original OP, but have subsequently engaged with the RS. It is this subsequent engagement that allows us to estimate a user's true preferences, \emph{retrospectively} correlate them with behavior in the OP, and generate realistic preference distributions. Together with the observable state, this comprises our \emph{user state generator}.

\begin{figure}
    \centering
\includegraphics[width=0.5\textwidth]{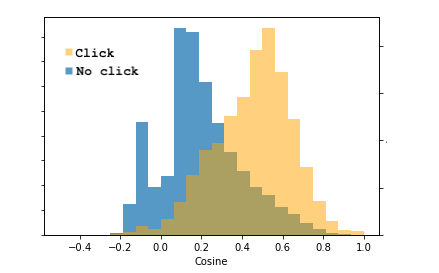}\,
\includegraphics[width=0.45\textwidth]{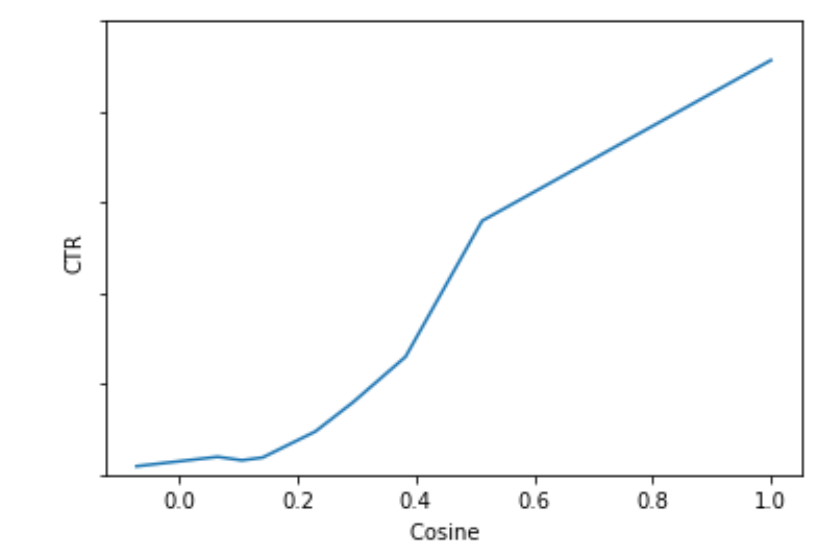}
    \caption{Histograms of cosine similarity between inferred user interests and selected/unselected artists (left). Artist click-through rate with respect to cosine similarity (right).}
    \label{fig:profile_analysis}
\end{figure}

Specifically, we augment OP session data with an ordered list $\{q^u_1, q^u_2, \ldots, q^u_K\}$ comprising the top-$K$ artists w.r.t.\ $u$'s post-onboarding music consumption time
during the first $W$ weeks of YouTube Music usage.%
\footnote{Initial RS recommendations will of course be correlated with onboarding selections and affect post-onboarding consumption. However, user searches and RS exploration will make this effect smaller as the horizon $W$ increases.}
We use this data below to train latent user-preference models.
Figure~\ref{fig:profile_analysis} (left) shows histograms of (embedding) similarity scores between users' top-$K$ post-onboarding artists (their ``true'' interests) and their selected (Click) or skipped (No click) artists during the OP, showing the greater correlation/similarity of actual music consumption with OP selections.
Figure \ref{fig:profile_analysis} (right) shows the click-through rate (CTR) during the OP as a function of the similarity of the query artist with post-onboarding artists: query artists that are more similar to post-OP artists are more likely to be selected during the OP.

To train the user state generator $P(C^u)$,
we can easily learn a categorical (joint) distribution over relevant observable state variables $C^u_o$ (e.g., geo, device) from our data.
However, directly modeling the joint distribution $P(C^u_\ell)$ of latent preferences (i.e., all combinations of inferred user interests) is intractable. Hence, we formulate our latent state model as a sequence generation or \emph{next-item prediction} problem \cite{yuan_etal:wsdm19} conditioned on the observable state $c_u$. Specifically, given a user's (post-OP) music consumption, we predict the next item $q^u_k$ given a user $u$'s past interaction history: $P(q^u_k | q^u_1, \ldots, q^u_{k-1})$. We factorize the joint distribution $P(C^u)$ as:
\begin{equation}
    P(C^u)\!=\! P(c^u, q^u_1, \ldots, q^u_K) = P(c^u) \!\prod_{k=1}^{K} P(q^u_k | c^u, q^u_1,\! \ldots,\! q^u_{k-1})
\end{equation}

\emph{Recurrent neural networks (RNNs)} \cite{hopfield_network} have proven effective for next-item prediction \cite{hidasi_etal:iclr16,liu_etal:icdm16,quadrana_etal:recsys17}. Thus, we use an RNN to model the temporal dependencies needed to generate inferred user interests $q^u_1, \ldots, q^u_K$. We apply a trainable multilayer perceptron (MLP) adapter to the one-hot encoding of $c^u$ and seed the initial state of the RNN  $h^u_0$ with the output. The RNN updates its hidden state $h^u_k$ and generates output $o^u_k$ as follows:
\begin{equation}
    (h^u_k, o^u_k) = \RNN(h^u_{k-1}, \phi(q^u_k), \delta_k)
\end{equation}
where $\phi(q^u_k)$ is the embedding of artist $q^u_k$ and $\delta_k$ is a one-hot encoding of the number of artists generated so far. We do not train the artist embedding layer, but instead reuse embeddings
from the PE module.
The next artist $q^u_{k+1}$ is then sampled (without replacement) using a standard \emph{multinomial logit} model \cite{louviere-et-al:statedchoice2000,chaptini2005use} given $o^u_k$:
\begin{equation}\label{eq:multinomial_logit}
P(q^u_{k+1} | o^u_k) = \frac{\exp(\phi(q^u_{k+1})^\top o^u_k)}{\sum_{q \in \calQ\setminus \{q^u_i\}_{i \leq k}} \exp(\phi(q)^\top o^u_k) + \exp(\phi_\varnothing^\top o^u_k)},
\end{equation}
To generate variable-length sequences, we introduce a fixed \emph{null artist} embedding $\phi_\varnothing$. We do not increase $k$ and change the input vector whenever the null artist is sampled. Including $\delta_k$ as an RNN input helps match the distribution of ``lengths'' of the number of artists in users' inferred interests to the observed data. 

Our latent state generator is implemented with a \emph{long short-term memory (LSTM)} architecture  \cite{lstm}.
By modeling generation of user interests $P(q^u_{k+1} | C^u, q^u_1, \ldots, q^u_k) = P(q^u_{k+1} | o^u_k)$ with an RNN and $P(C^u)$ as a categorical distribution, we can optimize parameters $\theta$ of the entire generator by maximum likelihood estimation (MLE):
\begin{equation}\label{eq:mle_cu}
    \max_\theta  \sum_{u \in \calU} \log P(C^u; \theta)
\end{equation}

\subsection{User Session Generator}

The second component of our model is a \emph{user session generator} that samples a user's action choice when presented with an artist, conditioned on their (static) state and their action choices thus far. User's attitudes evolve during the OP (e.g., degree of engagement/frustration with the OP; satisfaction with the information provided; try to steer the OPs future ``queries'').
These are part of the user's \emph{latent, dynamic state}: our model captures such dynamics implicitly to the extent they help generate realistic session behaviors.

To train the session generator, we assume a set of artists  $\calQ = \{q_1, \ldots, q_m\}$ eligible to be shown during PE. A session $S^u$ for $u\in\calU$ comprises a sequence of artists/queries $q^u_t$ and corresponding responses $r^u_t$: $S^u = \{(q^u_1, r^u_1), \ldots, (q^u_{|S^u|}, r^u_{|S^u|})\}$, where $r^u_t \in \{0, 1\}$ is the user response (skip or select) to $q^u_t$ at turn $t$. User termination occurs only at turn $t=|S^u|$. Define $S^u_{<t} = \{(q^u_1, r^u_1), \ldots, (q^u_{t-1}, r^u_{t-1})\}$ to be $u$'s sub-session prior to $t$. We note that this data format makes certain simplifying assumptions about the linear nature in which users inspect artists, which may not be precisely valid depending on the UI, since multiple artists are visible at once (slate size varies with device type).\footnote{We discuss our assumptions further in Appendix \ref{sec:app} Since the linear inspection assumption more closely approximates usage on mobile devices (vs., say, desktop), we confine our models and experiments to mobile platforms.}
We truncate each session $S^u$ at $300$ turns.

We formulate the user response model $P(r^u_t|C^u, S^u_{<t}, q^u_t)$ as contextual sequence generation,
using an RNN to encode \emph{(in-session) dynamic} user state and its dynamics. 
We first embed inferred user interests in $C^u$ via function $\phi$, to which we apply a trainable MLP to obtain encoded context $E(C^u)$ to personalize  session state, dynamics and output (user response) for any synthetic user $u$.
The input vector $x^u_t$ to the RNN at turn $t$ includes
$E(C^u)$, the embedding $\phi(q^u_t)$ of the artist/query, the embedding of selection response $\phi^r(r^u_t)$, the number $y_t$ of selections so far, and $t$. The latter two are critical for reproducing realistic distributions of the number of artist selections and session length.
The response embedding $\phi^r$ is trainable but, as above, we reuse artist embeddings $\{\phi(q)\}_{q \in \calQ}$ from PE.
The RNN hidden state $h^u_t$ at turn $t$ is updated as:
\begin{equation}
    (h^u_t, o^u_t) = \RNN(h^u_{t-1}, x^u_t)
\end{equation}
where $h^u_0 = E(C^u)$ and $o^u_t$ is the output of the RNN at turn $t$.

OP sessions $S^u$ can vary in length $|S^u|$ since $u$ determines when to terminate based on their latent state and the current artist; thus, the response model must also predict session continuation. We use an MLP with two heads and a sigmoid activation function $\sigma$ to implement selection response $r^u_t$
and session continuation $s^u_t$"
\begin{align}
\label{eq:response_dist_model}
    P(r^u_t=1|C^u, S^u_{<t}, q^u_t) = \sigma(\MLP(o^u_{t-1}, \Bar{x}^u_{t-1})) \cdot s^u_{t-1}, \\
\label{eq:continuation_dist_model}
    P(s^u_t=1|C^u, S^u_{<t}, q^u_t) = \sigma(\MLP(o^u_{t-1}, \Bar{x}^u_{t-1})) \cdot s^u_{t-1}.
\end{align}
We also pass $\Bar{x}^u_{t-1} = (E(C^u), y_{t-1}, t-1, \phi(q^u_t))$ to the MLP.
The response model does not predict a selection if the model predicts termination.
%
As above, the session generator is implemented using an LSTM, and we optimize its parameters by MLE.

\subsection{Transformer-based User Models}

In addition to RNNs, we experimented with \emph{transformers} \cite{vaswani_2017} to generate user context and sessions. Transformers perform similarly to RNNs, so we focus on RNNs in the following sections; but we describe the transformer for completeness.
Generating user context with a transformer is similar to the RNN-based approach. The input is the same sequence of vectors as used by the RNN.
Each input also includes a position embedding. We sample the $k+1$st artist using a multinomial logit (Eq.~\ref{eq:multinomial_logit}), where the $o_k^u$ is the transformer's output given the $k$th input. The sampled artist is appended to the input, and the process repeats to generate a user's preferences.

Generating a user session with a transformer is also similar to the RNN, though its input differs more;
the input sequence comprises: the sequence of inferred artists $(\phi(q_{1}^u), ..., \phi(q_K^u))$ is (static) state $C^u$, each concatenated with a learned embedding denoting user context; the artists queried so far, $(\phi(q^u_1), ..., \phi(q^u_{t-1}))$, each concatenated with its response $\phi^r(r^u)$ embedding; and the current query/artist embedding $\phi(q_t^u)$.
We sample a response and session continuation with Eqs.~\ref{eq:response_dist_model} and \ref{eq:continuation_dist_model}, using the transformer output $o^u_t$ given to the current artist/query input $\phi(q_t^u)$, along with $\Bar{x}_t^u$ as in the RNN.

\section{The Simulator}
\label{sec:sim}

We now describe our production-level simulator, which integrates our synthetic user models into the production stack to simulate users engaging in the real OP.
As discussed in Sec.~\ref{sec:pe_sim}, a controlled \emph{offline, synthetic user} experiment requires running the production RS with simulated users (vs.\ a mock RS). We create a simulation service that mocks user interactions, while
preserving production logging/serving. This poses challenges, including maintaining
separation between the simulated and the live system. To do so, we must ensure:
(a) The simulator runs against the full content-serving stack;
(b) Simulation does not interfere with live system function;
(c) Simulated-user data is separated from real-user data;
and (d) The production RS cannot distinguish simulated- and real-user data.
Separation of simulated and production data does not imply separation of the infrastructure accessing this data.
This too poses challenges since production infrastructure may not be designed to handle the usage patterns generated by a simulator (e.g., O(1B) existing users consuming content
might exercise the infrastructure differently than O(100K) new users onboarding at once). 

\vspace*{1.5mm}
\textbf{Simulating Music Onboarding.}
%
The user model interacts with the simulator through an API. This API provides the following capabilities:
(1) Navigation controls for ``looking'' at different artists in the slate; 
(2) Pre-seeding the (synthetic) user with music preferences to be reflected in the list of artists;
(3) Artist selection from the slate;
(4) Dynamic updating of the artist list presented to the synthetic user given any artist selection;
(5) Submission of all artist selections to be saved in the user profile to be consumed by the downstream RS; 
and (6) Navigation to the home page after submitting selections to see homepage results after users make their selections.

To ensure feature parity between the simulator and production RS, the simulator queries a front-end service which returns all elements to be rendered on the YouTube Music app. The front-end communicates with the RS back-end using the same content corpus and control flow as the live system.
For each synthetic user, the simulator creates a new \emph{test account} and requests the user-state generator to initialize simulated user data.
The front end includes the list of artists to display upon-load, taking into account the generated user state and additional artists to be dynamically loaded when the user selects an artist. The simulator parses this response and organizes artists linearly (see Sec.~\ref{sec:modeling}) to present to the user.

The user session generator selects an artist once the recommendations are served. The simulator responds to the front-end when an artist is selected, and the front-end adjusts the artist list accordingly.
Once the user session generator has finished selecting artists, it calls the API to submit the artists, which triggers a call to store these selections (i.e., our simulated user data) so as to affect downstream recommendations when the synthetic user is redirected to the YouTube Music homepage.

The interface to most current large-scale production infrastructure is request-based, where a request is typically associated with a single user. While underlying systems might batch or aggregate individual requests to exploit vectorized operations during model inference, not all systems support interaction in a batched manner. Thus, 
middleware may be necessary to bridge the two, and provide the scale needed to simulate many synthetic users.

\vspace*{1.5mm}
\textbf{Handling Simulated User Data.}
Storing artist selections by synthetic users must not change RS behavior  for real users. For instance, suppose the number of artist selections during the OP across some user segment changes which artists are recommended. Then synthetic artist selections must not increment those counts for real users.
Ensuring consistency and proper access control of (real and synthetic) user data for dependent systems
happens using a single \emph{user data serving service (UDSS)}.
In addition, we use a \emph{data overlay service (DOS)} which allows on-demand rewriting of synthetic data for the RS (not just the OP). 
In a live system, real user data is recorded by a production service and written to the UDSS. However, the simulator makes no such call; instead it calls the DOS to write the synthetic user history. Because the simulator runs in a controlled environment, we directly control synthetic user data.

There are two paths to the DOS, a write path and a read path. 
On the write path, DOS accepts a user specification as input, from which it creates simulated user data, which is stored  in its own storage system under each test account. The simulated user data generated by DOS fulfills any requirements needed by the RS consuming the data. On the read path, when UDSS receives a request, it recognizes whether is comes from our simulator. For such requests, UDSS makes an additional call to DOS to obtain its simulated user data, and either replaces the production data or merges with production data when generating a response.

Adding the DOS overlay 
only when reading user data
ensures no changes need to be made to upstream services.
Note that separation of simulated and production data may not imply a separation of infrastructure accessing simulated data. To provide a single point of access to upstream systems, DOS may need to interact with production infrastructure at a lower level than a typical ``client'' and therefore have to implement its own caching, batching, throttling, etc.\ to prevent disruption of production infrastructure.







\section{Experimental Results}
\label{sec:experiments}


We train RNN user models, using logged data of YouTube Music OP sessions, using the user-simulation development platform RecSim NG \cite{recsimNG:arxiv21}.
We deploy a simulation service for the OP, as described in Sec.~\ref{sec:sim}, that allows our simulated user models to interact with production infrastructure. We describe simulation results generated by our user models when interacting with both \emph{existing} and \emph{previously unseen} PE policies, focusing on validation of counterfactual robustness and policy optimization via offline simulation.

\subsection{Simulation with the Original Data Policy}
\label{sec:sim_with_orig}

\begin{figure}
    \centering
\includegraphics[width=0.45\textwidth]{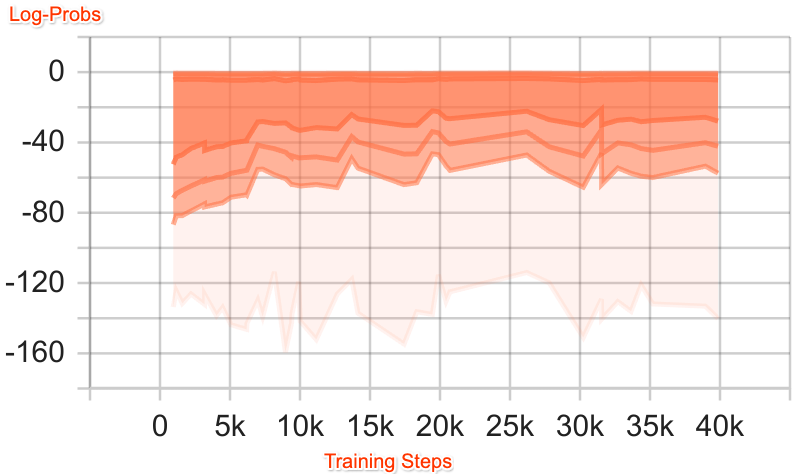}\,\includegraphics[width=0.45\textwidth]{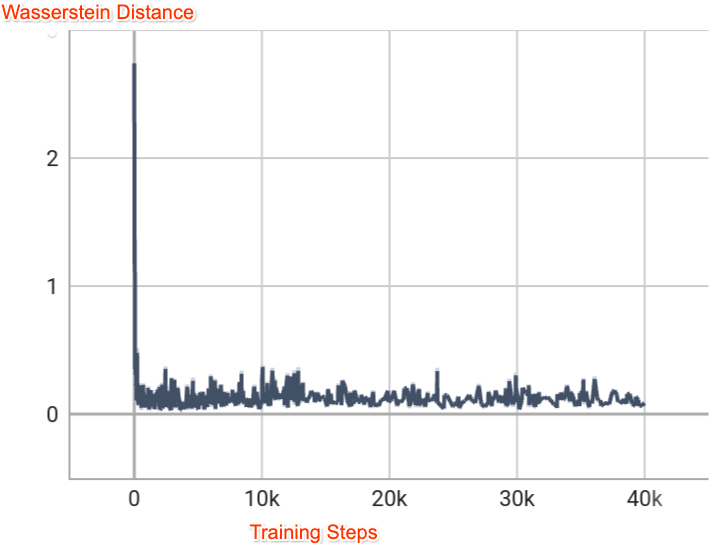}
    \caption{Convergence of training the user context generator: histogram of log-probabilities over number of training steps (left) and Wasserstein distance between the ground truth distribution of the number of artists in inferred user interests and the generated one (right).}
    \label{fig:profile_convergence}
\end{figure}

\begin{figure}
    \centering
\includegraphics[width=0.45\textwidth]{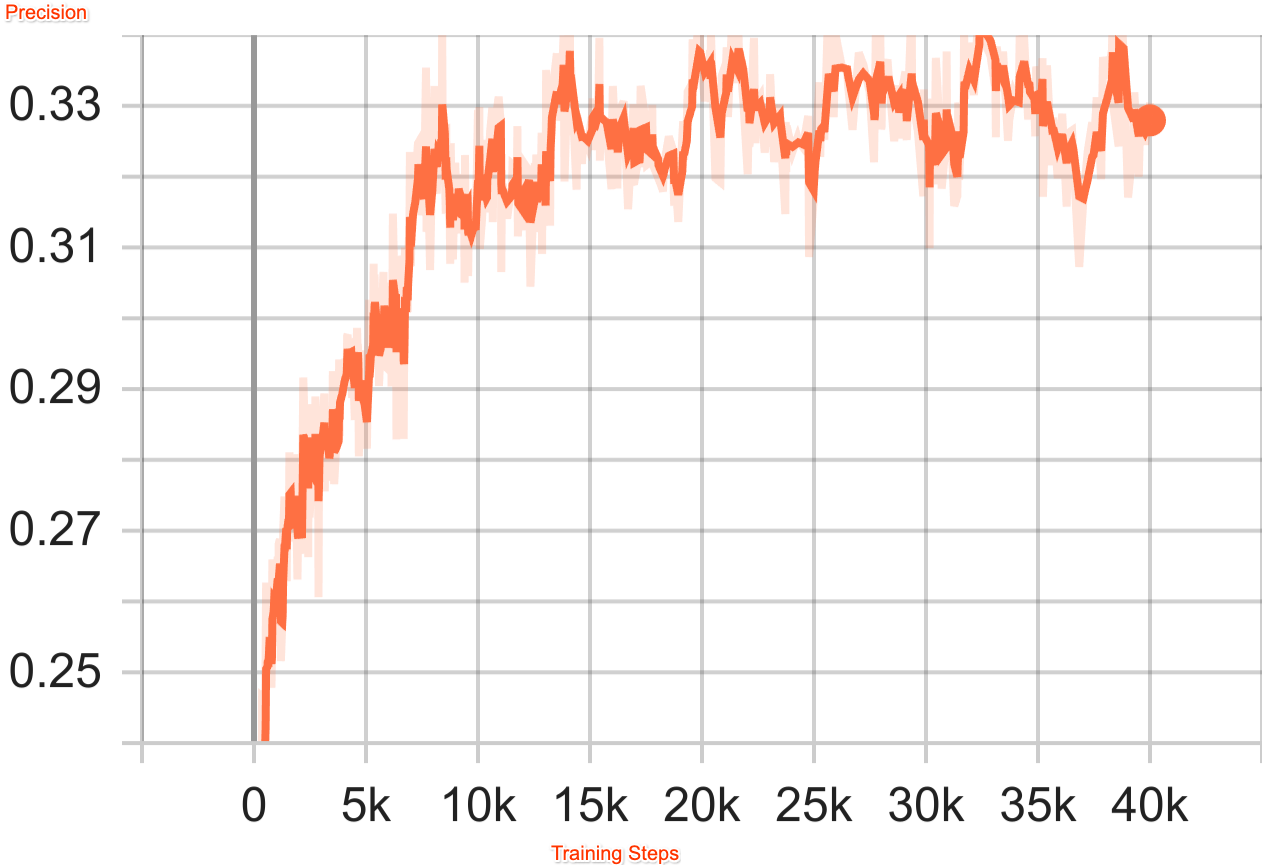}\,\includegraphics[width=0.45\textwidth]{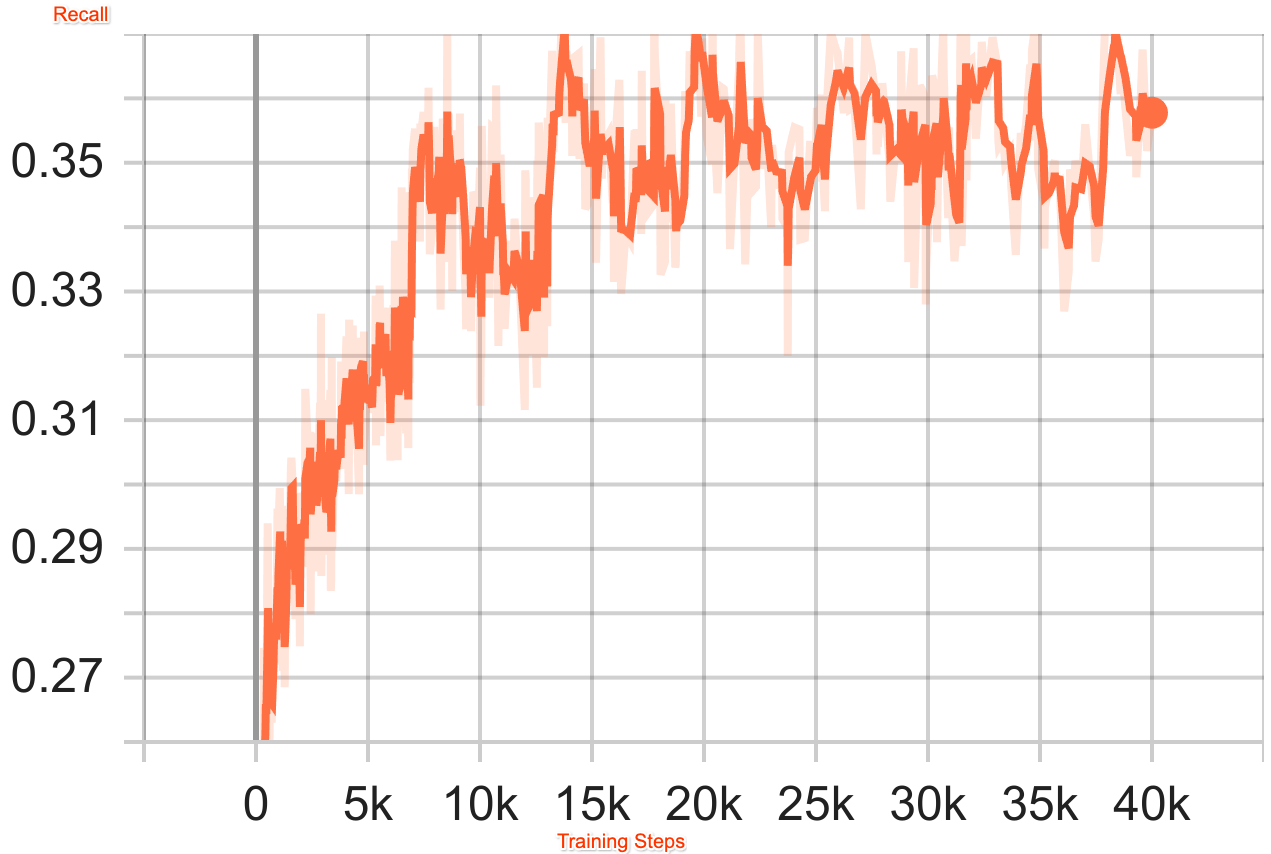}
    \caption{Precision (left) and recall (right) when comparing generated user interests with ground-truth user interests.}
    \label{fig:profile_precision_recall}
\end{figure}

Our first experiment exploits a deployed heuristic, well-performing OP policy, which serves as the behavior (i.e., data-generating) policy. Our user models are trained on this data, and we evaluate simulation quality by assessing whether it reproduces the distribution of key statistics observed in live (ground-truth) data. The user (static) state generator does not depend on the RS policy, so we simply compare the distribution of generated contexts with the ground-truth distribution. Figure~\ref{fig:profile_convergence} shows convergence of user state generator training. The histogram of log-probabilities (left) 
shows model fitness improves with training steps. The right plot shows the Wasserstein distance (over 4096 randomly sampled users) between the ground-truth and generated distributions of the number artists in each user's preferences.  We see that the model quickly reproduces a suitable distribution of user interest ``sizes.'' Precision and recall metrics in Figure~\ref{fig:profile_precision_recall} demonstrate the accuracy of generated- w.r.t.\ ground-truth preferences.

\begin{figure}
    \centering
\includegraphics[width=0.45\textwidth]{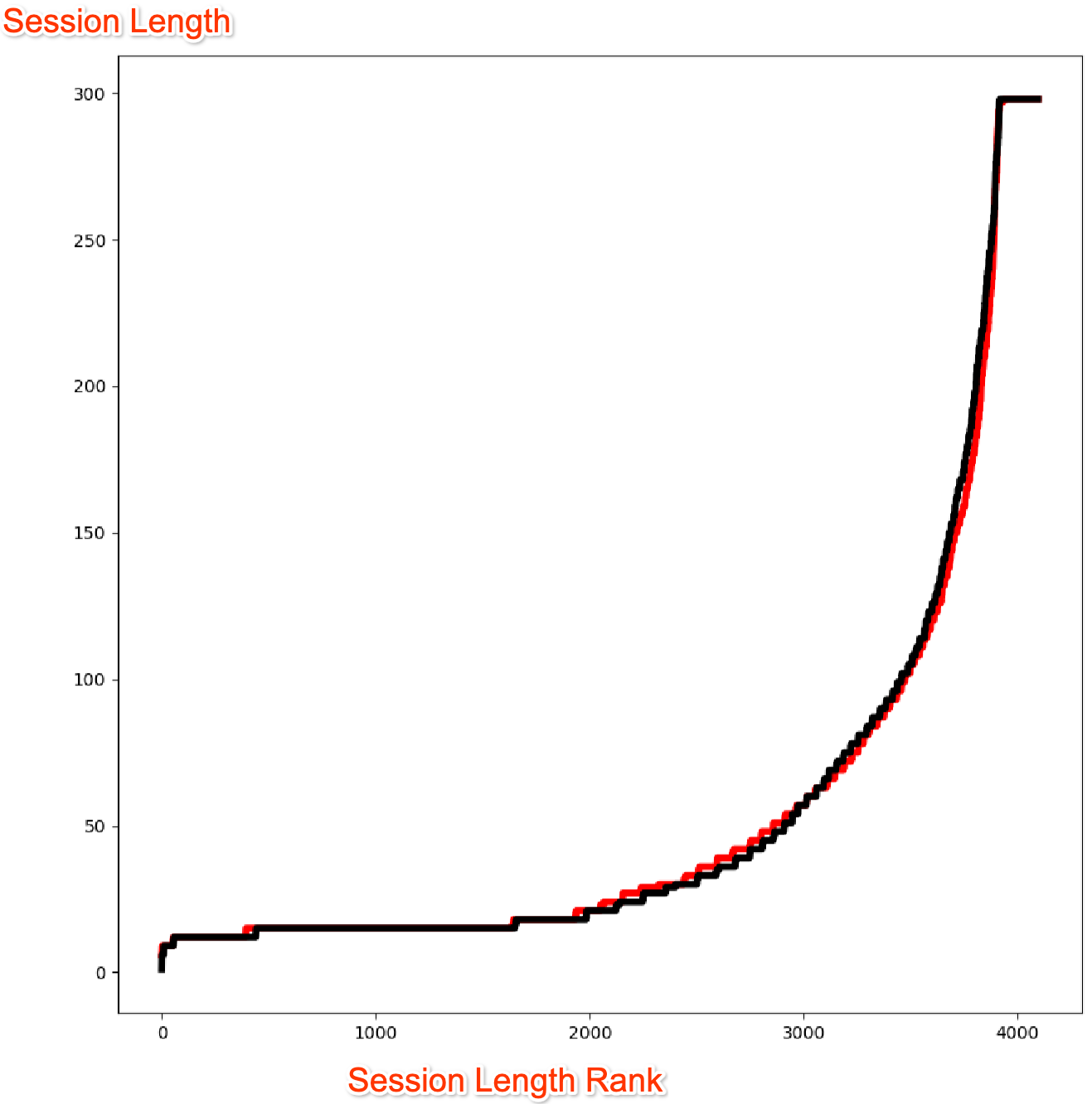}\,\includegraphics[width=0.45\textwidth]{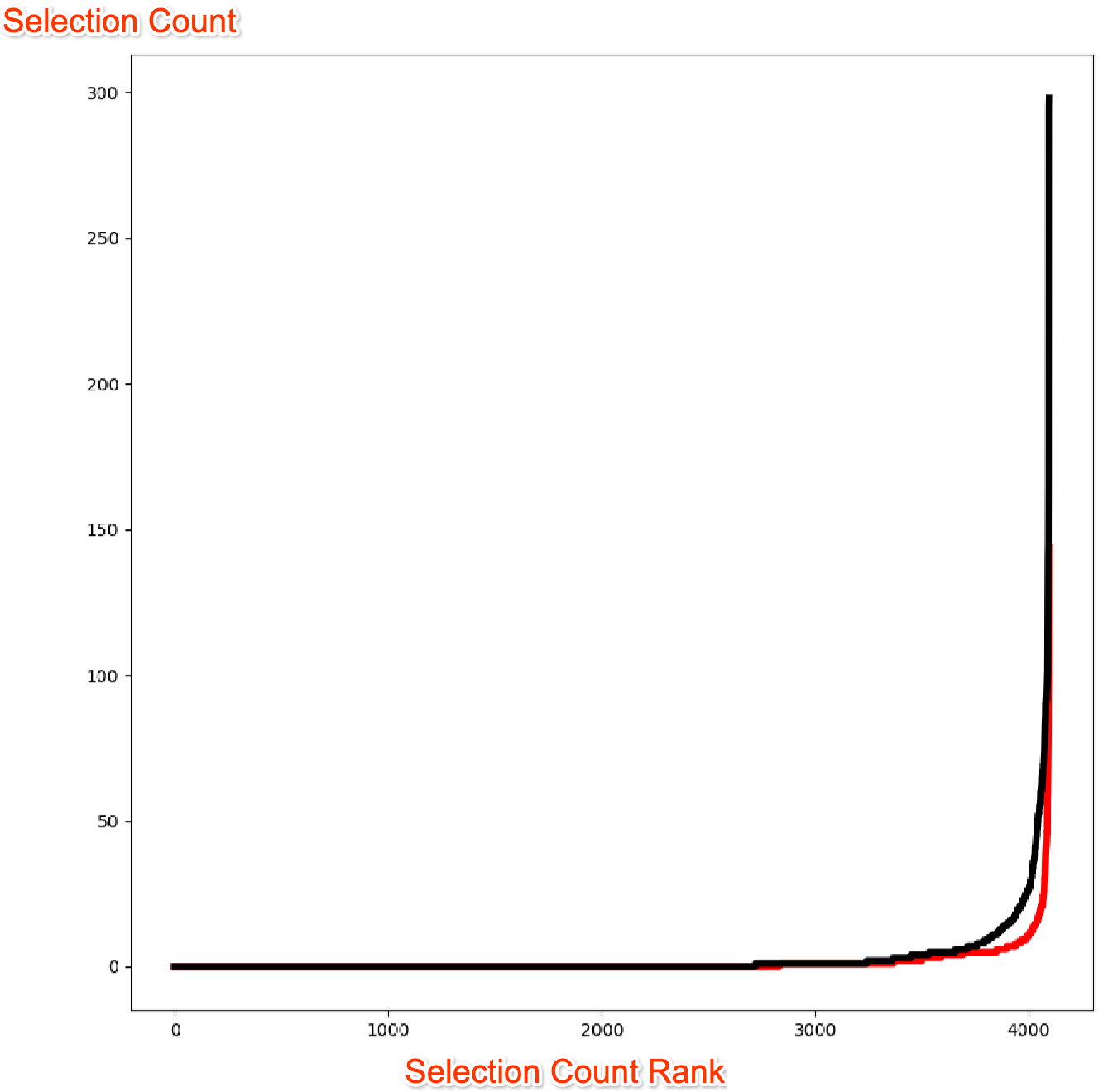}
    \caption{Cumulative distribution functions (CDFs) for both session lengths (left) and number of selections per session (right) by interacting with the data-generating policy where x-axis represents the rank in a batch of 4,096 sessions. Generated CDFs are red and observed CDFs in the log are black.}
    \label{fig:session_generator}
\end{figure}

For user session generation, we assess
the distribution of (i) the number of artist \emph{selections} per session and (ii) number of artist \emph{impressions} before termination (OP session length). Fig.~\ref{fig:session_generator} shows that the session generator reproduces the session-length distribution well: the cumulative distribution function (CDF) of generated lengths (red line) matches the observed CDF
(black line) well. It also reproduces the artist-selection count distribution relatively well (generated in red, observed/ground-truth in black).

\subsection{Simulation of New Policies}
\label{sec:sim_with_new}

Our end-to-end simulation results above with ``on-policy'' models (i.e., models interacting with behavior policies that generated their training data) show that the synthetic trajectories closely resemble those observed in logs. However, our goal of using simulation is to test (and ultimately optimize) \emph{new/unseen} OP PE policies requires that our user models exhibit a degree of counterfactual robustness. In other words, can they be used for off-policy evaluation to provide a reasonable assessment of (at least some) previously unseen policies.
We now turn to this assessment.

A critical user response during the OP is whether to select a presented query/artist (Eq.~\ref{eq:response_dist_model}). To this end, a new \emph{pCTR policy} was deployed in production: it exploits a model trained to predict the artist \emph{click-through rate (CTR)} (probability of selection) given the user's OP history so far,
and heuristically prioritizes artists with higher predicted CTR (pCTR) to increase onboarding selections. To validate the counterfactual robustness of our state and session generators---trained only on the \emph{previous} behavior policy's data---we simulate $200\,000$ synthetic users against the \emph{new pCTR policy} to forecast changes in both impressions and selections. We also ran LEs to gather the same online metrics with  $\tilde 134\,000$ real users.

Table~\ref{tab:pctr_us_android} provides the results of the new pCTR policy vs.\ the incumbent production policy (percentage change) with 95\% confidence intervals (CIs) on a (relatively large) traffic slice (one geo, one device type).
Employing such traffic slicing imposes difficulties for LEs, since the small test population (even for the largest slices) induces much wider CIs than simulated metrics---and shrinking the CI risks exposing significantly more users to a potentially poorly performing experimental policy. Thus, the ability to produce tighter CIs are a significant advantage of simulation. This leads to consistent overlap between online and simulated CIs. 
The LE and simulation both serve the common purpose of predicting the impact of the new policy if/when launched. In this case, the policy was launched, allowing us to compare the simulated metrics to the actual launch metrics---in Table~\ref{tab:pctr_us_android}, we compute post-launch metrics by sampling logs of $500\,000$ real OP users (this size provides much narrower CIs).\footnote{We assume metrics and user behaviors are stable over time in this pre-post analysis.} Moreover, while the LE is less helpful in predicting launch impact due to its wide CIs, our simulated predictions
are more closely aligned (informally, not statistically due to the wide LE CIs) with the real-world impact of the pCTR policy w.r.t \emph{actual post-launch metrics} obtained from 500,000 real users.

\begin{figure}
    \centering
\includegraphics[width=0.45\textwidth]{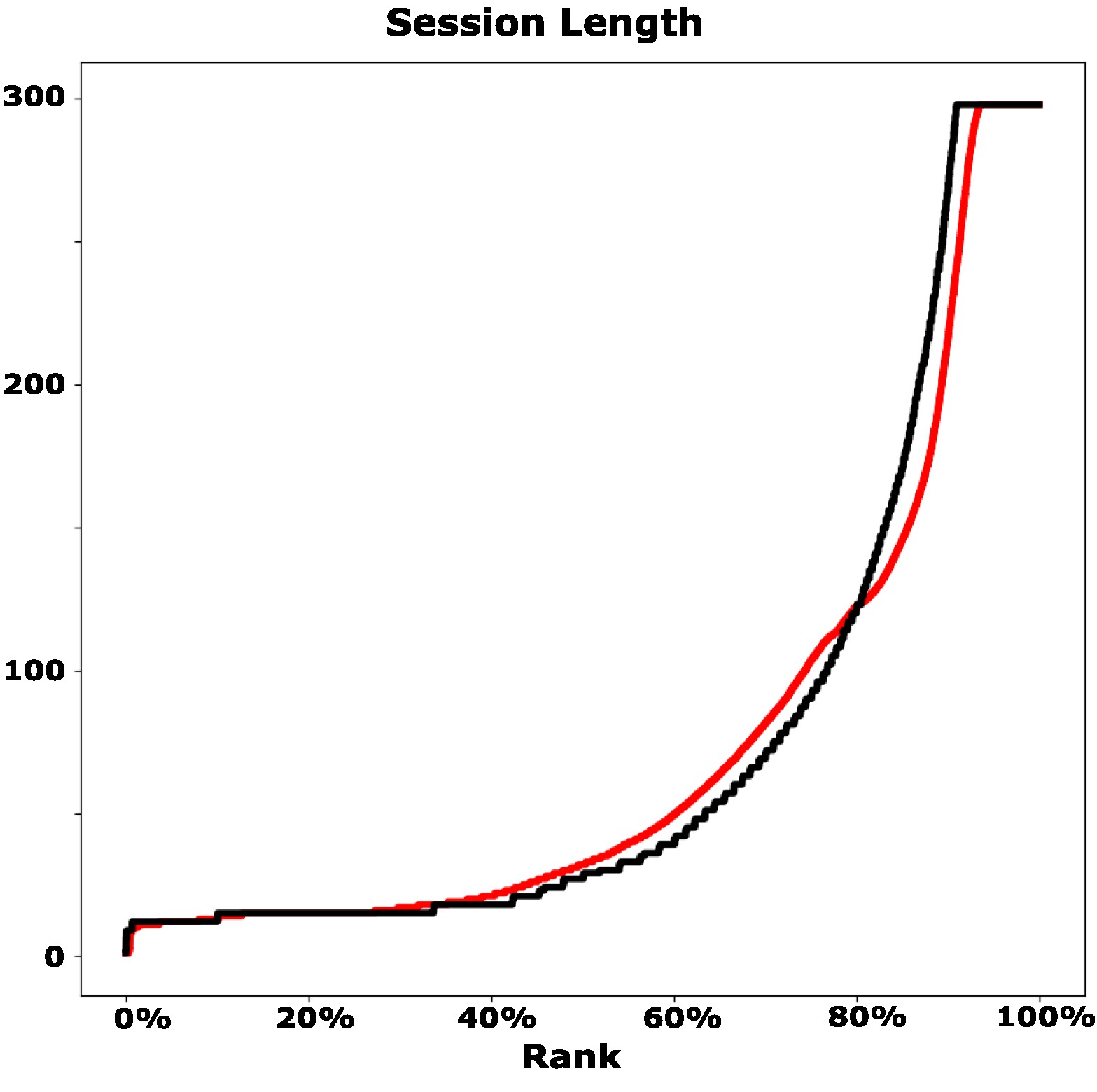}\,\includegraphics[width=0.45\textwidth]{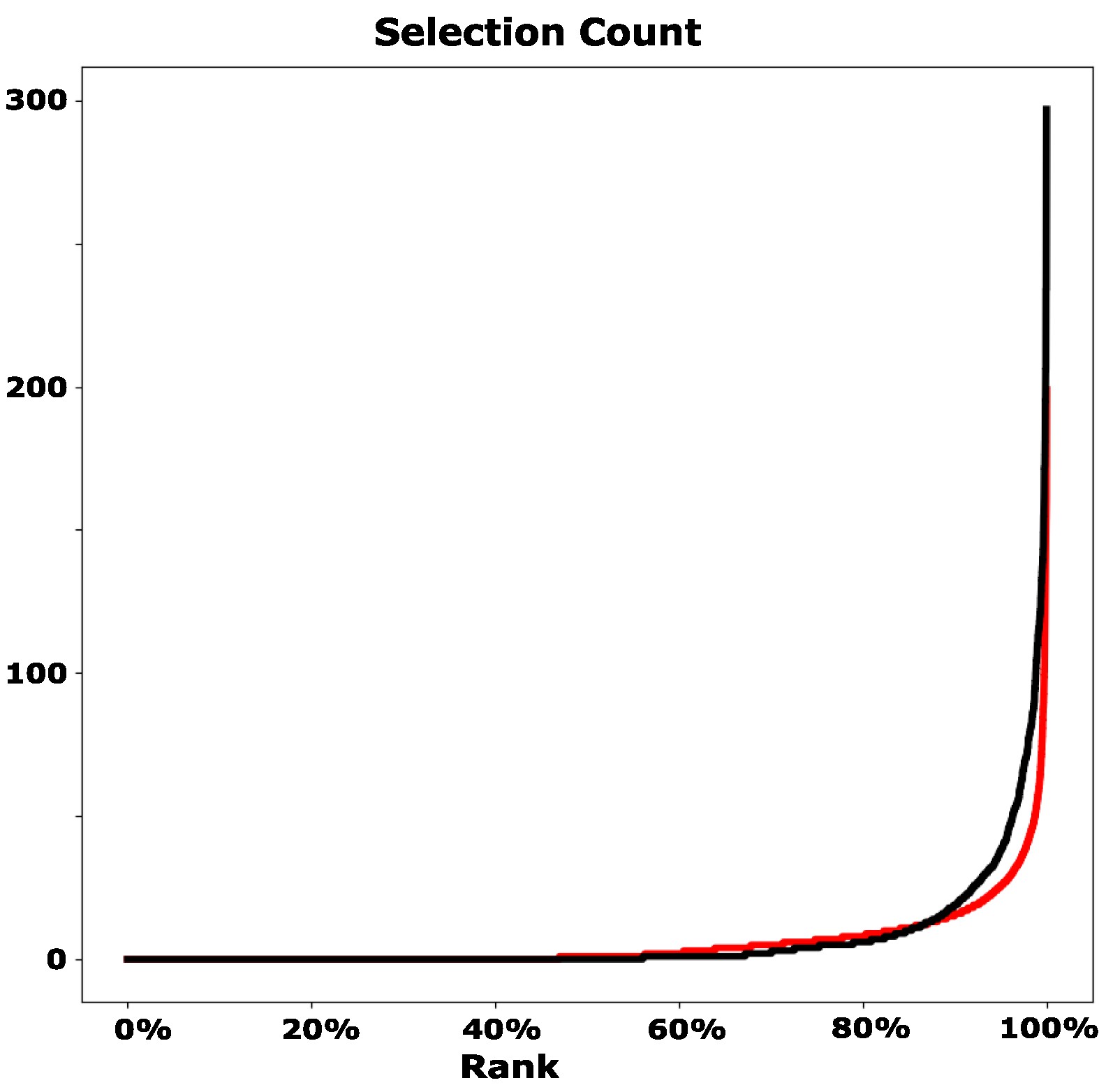}
    \caption{Cumulative distribution functions (CDFs) for both session length (left) and selection counts per session (right) by interacting with a new pCTR policy where x-axis represents the percentile rank. Generated CDFs are red and observed CDFs in post-launch log are black.}
    \label{fig:session_generator_new}
\end{figure}

Our simulation results also closely match the observed session length and artist-selection count distributions in post-launch logs, and faithfully reproduce the artist-selection count distribution,
as seen in Figure~\ref{fig:session_generator_new} (generated in red, observed in black).
Additionally, we evaluated AUCLoss of the selection prediction (Eq.~\ref{eq:response_dist_model}), which demonstrates that the user model has similar accuracy on both post-launch logs (0.760) and training data (0.763).

Naturally, this approach cannot, in general, \emph{guarantee} strong off-policy results. However, to the extent that (a) the user model is trained on a reasonable \emph{variety} of data/trajectories (due to policy variability and/or inherent stochasticity in user behaviors and contexts), and (b) the new policy is not radically different from the deployed policies used to generate training, a user simulator for new-policy evaluation can offer significant insights. Both of these conditions are met in our set up, which accounts for the strong performance. We next exploit this counterfactual robustness further to ``generate'' new policies.

\begin{table}
\centering
\begin{tabular}{|l||c|c|c|}
\hline
& \textbf{Users} & \textbf{Selections} & \textbf{Impressions} \\
\hline\hline
Live Exp. & \textasciitilde 134\,000 & -0.76\% [-5.05\%, 3.04\%] & -1.25\% [-3.57\%, 1.07\%] \\ \hline
Simulation & 200\,000 & +1.36\% [-0.11\%, 2.83\%] & -1.02\% [-1.74\%, -0.31\%] \\
\hline
Post-launch & 500\,000 & +1.17\% [-0.03\%, 2.36\%] & -0.78\% [-1.15\%, -0.41\%] \\
\hline
\end{tabular}
\caption{Comparison of Live Experiment, Simulated, and Post-launch Metrics of New pCTR Policy on a Traffic Slice.}
\label{tab:pctr_us_android}
\end{table}

\subsection{Offline Optimization}
\label{sec:sim_with_optimization}

Once validating model quality (Sec.~\ref{sec:sim_with_orig}) and partial counterfactual robustness (Sec.~\ref{sec:sim_with_new}),
we test whether we can use our user models to \emph{improve} the RS's policy \emph{purely offline} using simulation with synthetic users. The ability to do so reliably can significantly reduce both algorithm-development cycle time and the user-cost of running LEs to compare candidate algorithms.
In its simplest form, we use simulation to compare policies at a scale that is infeasible to test using LEs, and select the most promising policies for LE.
%
In the first test of this approach,
we simulated three \emph{new pCTR models}---these vary in the features used to predict an artist's selection probability and in their training data filters---and identified the best candidate: it has a \emph{simulated} CTR gain of $+1.15\% [-1.71\%, 4.01\%]$ over the original pCTR model.
This candidate was run in an LE with an observed CTR gain of $+2.06\% [0.20\%, 3.92\%]$; and it was subsequently launched (put into production).

\begin{table}
\centering
\begin{tabular}{|c||c|c|}
\hline
Trade-off $\lambda$ & Selections change & Impressions change \\
\hline \hline
0.001 & $0.07\% [-3.35\%, 3.49\%]$ & $0.75\% [-0.08\%, 1.59\%]$ \\
0.01 & $0.20\% [-3.16\%, 3.56\%]$ & $-0.04\% [-0.91\%, 0.82\%]$ \\ 
0.05 & $0.84\% [-2.33\%, 4.00\%]$ & $0.47\% [-0.23\%, 1.18\%]$ \\
0.2 & $2.35\% [-0.96\%, 5.67\%]$ & $0.50\% [-0.33\%, 1.33\%]$ \\
0.5 & $-0.48\% [-3.69\%, 2.72\%]$ & $-0.29\% [-1.06\%, 0.48\%]$ \\ 
1 & $0.24\% [-2.73\%, 3.21\%]$ & $0.19\% [-0.50\%, 0.88\%]$ \\ 
2 & $-0.78\% [-4.03\%, 2.46\%]$ & $0.75\% [0.02\%, 1.48\%]$ \\ 
5 & $-1.67\% [-5.22\%, 1.88\%]$ & $0.18\% [-0.78\%, 1.15\%]$ \\ 
10 & $-2.12\% [-5.62\%, 1.39\%]$ & $0.56\% [-0.28\%, 1.41\%]$ \\
\hline
\end{tabular}
\caption{Simulated Results (with 95\% CIs) when Tuning the Trade-off between pCTR and Coverage. }
\label{tab:offline_optimization}
\end{table}

We next conduct \emph{simulation-based experimentation} to test another \emph{class of PE/OP policies} for product launch. We compare versions of a new PE policy which vary a single parameter controlling the trade-off between an artist's pCTR and its \emph{coverage} of artist space. Specifically, each candidate query/artist $q$ is scored using
$\mathit{Score}(q) = \mathit{pCTR}(q) + \lambda \mathit{Coverage}(q)$: the first term is $q$'s pCTR and the second, $q$'s gain in coverage (see \citet{meshi2023preference}). A strong-performing production policy with $\lambda=0.1$ was launched.
We subsequently used simulation to tune the trade-off parameter $\lambda$, by enumerating nine values of $\lambda$ and testing if any generate better metrics than the production policy with $\lambda=0.1$. 
Table~\ref{tab:offline_optimization} presents our offline simulation (Sim.) results on all mobile traffic, comparing simulation results of our nine candidates with the simulated results of the original. Policies with
$\lambda=0.2$ and $\lambda=0.05$ provide the greatest selection gains, with only modest impression increases, but have wide CIs.
These results led to running LEs with policies using only these two values of $\lambda$ to test their actual improvements. To test if simulation can rule out poor policies, we also run
$\lambda=0.001$ in LE (it was effectively neutral w.r.t.\ selections).

To test the hypothesis that large-scale simulation can more accurately order policies w.r.t.\ their launch metrics than small LEs (whose size may be constrained for ``cost'' reasons), we run one large ``launch LE'' over a specific period with the three policies ($\lambda\in \{0.001, 0.05, 0.2\}$) to measure their impact on the number of artist selections in an OP session.\footnote{We focus on artist selections here, but in Appendix \ref{sec:additional_results} we show similar results for impressions (which are somewhat noisier).} We then partition the ``launch'' data by looking at three subperiods, treating each as if it were a smaller scale ``true LE.'' We assess the policy ordering w.r.t.\ each ``true LE'' to assess consistency with the launch data; we evaluate our simulation similarly.\footnote{We note that the ``LEs'' and ``Launch'' metrics will be more correlated in our design than they would be in practice, since each LE is in fact a one-third sample of the Launch traffic.} Our results in Table~\ref{tab:LEvsLaunch} show the variation across the three small ``true LEs,'' none of which order the policies in the same way, and only one of which conforms to the ``launch;'' while the simulation ordering predicts (informally) that of the launch. While this test is imperfect (since we do not want to incur the cost of ``launch-scale'' LEs on multiple policies), it suggests that simulation can be used to effectively order policies offline.

\begin{table}
\centering
\begin{tabular}{|c||c|c|c|c|}
\hline
           & Sels. $\lambda=0.001$ & Sels. $\lambda=0.05$ & Sels. $\lambda=0.2$ & Ordering \\
\hline \hline
LE 1 & $0.71 \pm 3.35\% $ & $1.60 \pm 1.91\% $ & $1.17 \pm 3.81\% $ & $0.05 \!\!>\!\! 0.2 \!\!>\!\! 0.001$ \\
LE 2 & $1.62 \pm 3.40\% $ & $0.54 \pm 2.96\% $ & $0.52 \pm 3.38\% $ & $0.001 \!\!>\!\! 0.05 \!\!>\!\! 0.2$ \\ 
LE 3 & $-2.11 \pm 2.50\% $ & $0.07 \pm 3.02\% $ & $0.61 \pm 3.36\% $ & $0.2 \!\!>\!\! 0.05 \!\!>\!\! 0.001$ \\
Launch & $-0.12 \pm 1.88\% $ & $0.75 \pm 1.22\% $ & $0.79 \pm 1.98\% $ & $0.2 \!\!>\!\! 0.05 \!\!>\!\! 0.001$ \\
Simul. & $0.07 \pm 3.42\% $ & $0.84 \pm 3.16\% $ & $2.35 \pm 3.32\% $ & $0.2 \!\!>\!\! 0.05 \!\!>\!\! 0.001$ \\ \hline
\end{tabular}
\caption{Selection Change (95\% CIs) for 3 Policies, and Induced Policy Ordering: LEs vs.\ ``Launch'' vs. Simulation.}
\label{tab:LEvsLaunch}
\end{table}

\section{Simulation-based RS Development}
\label{sec:production}

We have demonstrated above the use of
user
simulation for some of the evaluation needed during RS development typically handled using costly LEs. This idea, taken to the extreme, paints a picture of a development and testing workflow largely free of LEs, where most evaluation takes place \emph{in silico}.
However, deploying such a workflow across a broad selection of RS tasks, beyond onboarding, faces
many methodological challenges before it can be deployed as a production-ready technology, some of which we outline below (we also speculate on potential solutions).

{\bf Counterfactual robustness.} Our methodology relies on the counterfactual robustness of the user model one develops; that is, it exploits a user model, trained on user behaviors induced by a (small) set of deployed \emph{behavior policies}, to evaluate the performance of a new policy (w.r.t.\ relevant metrics) not yet encountered by users. Without special accommodation, a model so-trained
may overfit 
to the behavior policies. Promising solutions to this may be found in \emph{causal modeling} \cite{pearl_book,elements_of_causal_inference}, 
though approaches that enforce constraints on model structure (e.g., mechanistic plausibility) might
limit the use of high-capacity models (e.g., transformers).
Notions like \emph{$mz$-transportability} \cite{mz_transportability} may prove useful, by
forcing a model to learn patterns common to multiple behavior policies,
so that what is learned is (ideally) independent of the RS policy.
Given that production RSs deploy numerous policies (and LEs) over time,
the natural variation in behavior policies and data distributions can be harnessed
for effective
model training. How to select a minimal useful set
remains an open question. 

{\bf Exploration and exploitation.} From the point of view of the RS lifecycle, simulated A/B testing is effectively a \emph{bandit problem} \cite{lai85asymptotically,auer02finitetime,lattimore19bandit}: policies are arms whose value/metrics
are predicted by the simulator. Since all models are imperfect, the value of a policy may be either overestimated or underestimated. Overestimation may be less problematic than underestimation in the long-run---if a policy's value is overestimated, a policy will ``graduate'' to an A/B test in an LE, where its true value will be revealed (at a cost). Ideally, data generated from such an LE is used to improve the simulation model as well, so that model deficiencies inducing overestimation can be rectified.  Underestimation may be more troublesome, since good policies may be discarded prematurely, leaving value/progress ``on the table'' and biasing the development process toward what the simulator is able to accurately predict. Avoiding this requires live A/B testing of apparently suboptimal policies, at a rate ideally determined by calibrated uncertainty estimates (also taking environment non-stationarity into account). From a Bayesian perspective, we can think of the uncertainty in the predicted metrics as the product of the uncertainty in the model parameters and the uncertainty of the prediction conditioned on model parameters. The latter can be easy to narrow down by simply increasing the synthetic sample size. However, this does not help with the former uncertainty, which needs to be addressed with proper uncertainty quantification methods. 

{\bf Reliability.} If the aforementioned uncertainty estimates are available, a reliable measure of model risk can be used to judge whether simulation results can be trusted. Since this can be challenging, as noted above, there is significant value in developing proxy heuristics that indicate whether simulation results are trustworthy before an LE is carried out. One proxy is to measure how different a test policy is from the behavior policies used to train the model. A smoothness argument suggests a model prediction is more reliable if the test policy is closer to the training policies. Closeness should be measured, however, in the latent space of the model, which may lead to non-trivial estimation problems. 

Another potential heuristic, applicable to models with latent state, is to measure the \emph{predictiveness} of its latent representation: if two partial sequences of interactions are compressed into the same latent representation, then the distribution of sequence continuations conditioned on one should be similar to that conditioned on the other. Predictiveness is a \emph{sufficient} (but not necessary) condition for reliability. Ideas for measuring predictiveness can be found in the literature on predictive state representations \cite{littman_psr}.

\section{Conclusion}
\label{sec:conclusion}
In this paper we demonstrated that simulation can play a key role in the evaluation and optimization of RS policies by reducing the need for costly live experiments.
We focused on the new user onboarding process in YouTube Music and showed how realistic user models can be learned from logged onboarding sessions and post-onboarding consumption.
We described a simulation platform that allows the integration of our synthetic user models will full production infrastructure with suitable separation of live and simulated data, and demonstrated the potential of our approach with empirical studies of YouTube Music onboarding. Finally, we outlined some avenues for future research in the promising area and wider adoption of simulation for RS algorithm development and evaluation.

\bibliographystyle{ACM-Reference-Format}
\balance
\bibliography{long,refs}


\begin{thebibliography}{55}


\ifx \showCODEN    \undefined \def \showCODEN     #1{\unskip}     \fi
\ifx \showDOI      \undefined \def \showDOI       #1{#1}\fi
\ifx \showISBNx    \undefined \def \showISBNx     #1{\unskip}     \fi
\ifx \showISBNxiii \undefined \def \showISBNxiii  #1{\unskip}     \fi
\ifx \showISSN     \undefined \def \showISSN      #1{\unskip}     \fi
\ifx \showLCCN     \undefined \def \showLCCN      #1{\unskip}     \fi
\ifx \shownote     \undefined \def \shownote      #1{#1}          \fi
\ifx \showarticletitle \undefined \def \showarticletitle #1{#1}   \fi
\ifx \showURL      \undefined \def \showURL       {\relax}        \fi
\providecommand\bibfield[2]{#2}
\providecommand\bibinfo[2]{#2}
\providecommand\natexlab[1]{#1}
\providecommand\showeprint[2][]{arXiv:#2}

\bibitem[Abel et~al\mbox{.}(2011)]%
        {Abel2011}
\bibfield{author}{\bibinfo{person}{Fabian Abel}, \bibinfo{person}{Qi Gao},
  \bibinfo{person}{Geert-Jan Houben}, {and} \bibinfo{person}{Ke Tao}.}
  \bibinfo{year}{2011}\natexlab{}.
\newblock \showarticletitle{Analyzing user modeling on {Twitter} for
  personalized news recommendations}. In \bibinfo{booktitle}{\emph{User
  Modeling, Adaption and Personalization}}.
\newblock


\bibitem[Ahlgren et~al\mbox{.}(2021)]%
        {ahlgren-facebookSim:ease21}
\bibfield{author}{\bibinfo{person}{John Ahlgren}, \bibinfo{person}{Kinga
  Bojarczuk}, \bibinfo{person}{Sophia Drossopoulou}, \bibinfo{person}{Inna
  Dvortsova}, \bibinfo{person}{Johann George}, \bibinfo{person}{Natalija
  Gucevska}, \bibinfo{person}{Mark Harman}, \bibinfo{person}{Maria Lomeli},
  \bibinfo{person}{Simon~Mark Lucas}, \bibinfo{person}{Erik Meijer},
  \bibinfo{person}{Steve Omohundro}, \bibinfo{person}{Rubmary Rojas},
  \bibinfo{person}{Silvia Sapora}, \bibinfo{person}{Jie~M. Zhang}, {and}
  \bibinfo{person}{Norm Zhou}.} \bibinfo{year}{2021}\natexlab{}.
\newblock \showarticletitle{Facebook’s Cyber--Cyber and Cyber--Physical
  Digital Twins}. In \bibinfo{booktitle}{\emph{Proceedings of the International
  Conference on Evaluation and Assessment in Software Engineering (EASE-21)}}.
  \bibinfo{pages}{1--9}.
\newblock


\bibitem[Auer et~al\mbox{.}(2002)]%
        {auer02finitetime}
\bibfield{author}{\bibinfo{person}{Peter Auer}, \bibinfo{person}{Nicolo
  Cesa-Bianchi}, {and} \bibinfo{person}{Paul Fischer}.}
  \bibinfo{year}{2002}\natexlab{}.
\newblock \showarticletitle{Finite-time Analysis of the Multiarmed Bandit
  Problem}.
\newblock \bibinfo{journal}{\emph{Machine Learning}}  \bibinfo{volume}{47}
  (\bibinfo{year}{2002}), \bibinfo{pages}{235--256}.
\newblock


\bibitem[Balog and Radlinski(2020)]%
        {balog_etal:sigir20}
\bibfield{author}{\bibinfo{person}{Krisztian Balog} {and}
  \bibinfo{person}{Filip Radlinski}.} \bibinfo{year}{2020}\natexlab{}.
\newblock \showarticletitle{Measuring Recommendation Explanation Quality: The
  Conflicting Goals of Explanations}. In \bibinfo{booktitle}{\emph{Proceedings
  of the 43rd International {ACM} {SIGIR} conference on research and
  development in Information Retrieval, {SIGIR} 2020, Virtual Event, China,
  July 25-30, 2020}}. \bibinfo{pages}{329--338}.
\newblock


\bibitem[Bansal et~al\mbox{.}(2019)]%
        {bansal_etal:hcomp19}
\bibfield{author}{\bibinfo{person}{Gagan Bansal}, \bibinfo{person}{Besmira
  Nushi}, \bibinfo{person}{Ece Kamar}, \bibinfo{person}{Walter~S. Lasecki},
  \bibinfo{person}{Daniel~S. Weld}, {and} \bibinfo{person}{Eric Horvitz}.}
  \bibinfo{year}{2019}\natexlab{}.
\newblock \showarticletitle{Beyond Accuracy: The Role of Mental Models in
  Human-AI Team Performance}. In \bibinfo{booktitle}{\emph{Proceedings of the
  Seventh {AAAI} Conference on Human Computation and Crowdsourcing, {HCOMP}
  2019, Stevenson, WA, USA, October 28-30, 2019}}. \bibinfo{pages}{2--11}.
\newblock


\bibitem[Bareinboim and Pearl(2014)]%
        {mz_transportability}
\bibfield{author}{\bibinfo{person}{Elias Bareinboim} {and}
  \bibinfo{person}{Judea Pearl}.} \bibinfo{year}{2014}\natexlab{}.
\newblock \showarticletitle{Transportability from Multiple Environments with
  Limited Experiments: Completeness Results}. In
  \bibinfo{booktitle}{\emph{Advances in Neural Information Processing Systems
  27: Annual Conference on Neural Information Processing Systems 2014, December
  8-13 2014, Montreal, Quebec, Canada}}. \bibinfo{pages}{280--288}.
\newblock


\bibitem[Beutel et~al\mbox{.}(2018)]%
        {beutel_etal:wsdm18}
\bibfield{author}{\bibinfo{person}{Alex Beutel}, \bibinfo{person}{Paul
  Covington}, \bibinfo{person}{Sagar Jain}, \bibinfo{person}{Can Xu},
  \bibinfo{person}{Jia Li}, \bibinfo{person}{Vince Gatto}, {and}
  \bibinfo{person}{Ed~H. Chi}.} \bibinfo{year}{2018}\natexlab{}.
\newblock \showarticletitle{Latent Cross: Making Use of Context in Recurrent
  Recommender Systems}. In \bibinfo{booktitle}{\emph{Proceedings of the
  Eleventh {ACM} International Conference on Web Search and Data Mining
  (WSDM-18)}}. \bibinfo{address}{Marina Del Rey, CA}, \bibinfo{pages}{46--54}.
\newblock


\bibitem[Bobadilla et~al\mbox{.}(2012)]%
        {bobadilla2012}
\bibfield{author}{\bibinfo{person}{Jes{\'u}S Bobadilla},
  \bibinfo{person}{Fernando Ortega}, \bibinfo{person}{Antonio Hernando}, {and}
  \bibinfo{person}{Jes{\'u}s Bernal}.} \bibinfo{year}{2012}\natexlab{}.
\newblock \showarticletitle{A collaborative filtering approach to mitigate the
  new user cold start problem}.
\newblock \bibinfo{journal}{\emph{Knowledge-based systems}}
  \bibinfo{volume}{26} (\bibinfo{year}{2012}), \bibinfo{pages}{225--238}.
\newblock


\bibitem[Boutilier(2002)]%
        {preference:aaai02}
\bibfield{author}{\bibinfo{person}{Craig Boutilier}.}
  \bibinfo{year}{2002}\natexlab{}.
\newblock \showarticletitle{A {POMDP} Formulation of Preference Elicitation
  Problems}. In \bibinfo{booktitle}{\emph{Proceedings of the Eighteenth
  National Conference on Artificial Intelligence (AAAI-02)}}.
  \bibinfo{address}{Edmonton}, \bibinfo{pages}{239--246}.
\newblock


\bibitem[Chajewska et~al\mbox{.}(2000)]%
        {chajewska2000making}
\bibfield{author}{\bibinfo{person}{Urszula Chajewska}, \bibinfo{person}{Daphne
  Koller}, {and} \bibinfo{person}{Ronald Parr}.}
  \bibinfo{year}{2000}\natexlab{}.
\newblock \showarticletitle{Making Rational Decisions Using Adaptive Utility
  Elicitation}. In \bibinfo{booktitle}{\emph{Proceedings of the Seventeenth
  National Conference on Artificial Intelligence (AAAI-00)}}.
  \bibinfo{address}{Austin, TX}, \bibinfo{pages}{363--369}.
\newblock


\bibitem[Chaptini(2005)]%
        {chaptini2005use}
\bibfield{author}{\bibinfo{person}{Bassam~H. Chaptini}.}
  \bibinfo{year}{2005}\natexlab{}.
\newblock \emph{\bibinfo{title}{Use of Discrete Choice Models with Recommender
  Systems}}.
\newblock \bibinfo{thesistype}{Ph.\,D. Dissertation}.
  \bibinfo{school}{Massachusetts Institute of Technology},
  \bibinfo{address}{Cambridge, MA}.
\newblock


\bibitem[Chen et~al\mbox{.}(2018)]%
        {chen_etal:2018top}
\bibfield{author}{\bibinfo{person}{Minmin Chen}, \bibinfo{person}{Alex Beutel},
  \bibinfo{person}{Paul Covington}, \bibinfo{person}{Sagar Jain},
  \bibinfo{person}{Francois Belletti}, {and} \bibinfo{person}{Ed Chi}.}
  \bibinfo{year}{2018}\natexlab{}.
\newblock \showarticletitle{Top-K Off-Policy Correction for a {REINFORCE}
  Recommender System}. In \bibinfo{booktitle}{\emph{12th {ACM} International
  Conference on Web Search and Data Mining (WSDM-19)}}.
  \bibinfo{address}{Melbourne, Australia}, \bibinfo{pages}{456--464}.
\newblock


\bibitem[Covington et~al\mbox{.}(2016a)]%
        {Covington2016}
\bibfield{author}{\bibinfo{person}{Paul Covington}, \bibinfo{person}{Jay
  Adams}, {and} \bibinfo{person}{Emre Sargin}.}
  \bibinfo{year}{2016}\natexlab{a}.
\newblock \showarticletitle{Deep neural networks for {YouTube}
  recommendations}. In \bibinfo{booktitle}{\emph{Proceedings of the 10th ACM
  Conference on Recommender Systems}}.
\newblock


\bibitem[Covington et~al\mbox{.}(2016b)]%
        {covington_etal:recsys16}
\bibfield{author}{\bibinfo{person}{Paul Covington}, \bibinfo{person}{Jay
  Adams}, {and} \bibinfo{person}{Emre Sargin}.}
  \bibinfo{year}{2016}\natexlab{b}.
\newblock \showarticletitle{Deep Neural Networks for YouTube Recommendations}.
  In \bibinfo{booktitle}{\emph{Proceedings of the 10th {ACM} Conference on
  Recommender Systems, Boston, MA, USA, September 15-19, 2016}}.
  \bibinfo{pages}{191--198}.
\newblock


\bibitem[Craswell et~al\mbox{.}(2008)]%
        {cascade_model}
\bibfield{author}{\bibinfo{person}{Nick Craswell}, \bibinfo{person}{Onno
  Zoeter}, \bibinfo{person}{Michael~J. Taylor}, {and} \bibinfo{person}{Bill
  Ramsey}.} \bibinfo{year}{2008}\natexlab{}.
\newblock \showarticletitle{An experimental comparison of click position-bias
  models}. In \bibinfo{booktitle}{\emph{Proceedings of the International
  Conference on Web Search and Web Data Mining, {WSDM} 2008, Palo Alto,
  California, USA, February 11-12, 2008}}. \bibinfo{pages}{87--94}.
\newblock


\bibitem[Dud{\'{\i}}k et~al\mbox{.}(2011)]%
        {doubly_robust}
\bibfield{author}{\bibinfo{person}{Miroslav Dud{\'{\i}}k},
  \bibinfo{person}{John Langford}, {and} \bibinfo{person}{Lihong Li}.}
  \bibinfo{year}{2011}\natexlab{}.
\newblock \showarticletitle{Doubly Robust Policy Evaluation and Learning}. In
  \bibinfo{booktitle}{\emph{Proceedings of the Twenty-eighth International
  Conference on Machine Learning (ICML-11)}}. \bibinfo{address}{Bellevue, WA},
  \bibinfo{pages}{1097--1104}.
\newblock


\bibitem[Hallinan and Striphas(2016)]%
        {Hallinan2016}
\bibfield{author}{\bibinfo{person}{Blake Hallinan} {and} \bibinfo{person}{Ted
  Striphas}.} \bibinfo{year}{2016}\natexlab{}.
\newblock \showarticletitle{Recommended for you: The {Netflix} Prize and the
  production of algorithmic culture}.
\newblock \bibinfo{journal}{\emph{New Media \& Society}}  \bibinfo{volume}{18}
  (\bibinfo{year}{2016}), \bibinfo{pages}{117--137}.
\newblock


\bibitem[Hidasi et~al\mbox{.}(2016)]%
        {hidasi_etal:iclr16}
\bibfield{author}{\bibinfo{person}{Bal{\'{a}}zs Hidasi},
  \bibinfo{person}{Alexandros Karatzoglou}, \bibinfo{person}{Linas Baltrunas},
  {and} \bibinfo{person}{Domonkos Tikk}.} \bibinfo{year}{2016}\natexlab{}.
\newblock \showarticletitle{Session-based Recommendations with Recurrent Neural
  Networks}. In \bibinfo{booktitle}{\emph{Proceedings of the Fourth
  International Conference on Learning Representations (ICLR-16)}}.
  \bibinfo{address}{San Juan, Puerto Rico}.
\newblock


\bibitem[Hochreiter and Schmidhuber(1997)]%
        {lstm}
\bibfield{author}{\bibinfo{person}{Sepp Hochreiter} {and}
  \bibinfo{person}{J{\"{u}}rgen Schmidhuber}.} \bibinfo{year}{1997}\natexlab{}.
\newblock \showarticletitle{Long Short-Term Memory}.
\newblock \bibinfo{journal}{\emph{Neural Computation}} \bibinfo{volume}{9},
  \bibinfo{number}{8} (\bibinfo{year}{1997}), \bibinfo{pages}{1735--1780}.
\newblock


\bibitem[Hopfield(1982)]%
        {hopfield_network}
\bibfield{author}{\bibinfo{person}{J.~J. Hopfield}.}
  \bibinfo{year}{1982}\natexlab{}.
\newblock \showarticletitle{Neural Networks and Physical Systems with Emergent
  Collective Computational Abilities}.
\newblock \bibinfo{journal}{\emph{Proceedings of the National Academy of
  Sciences}} \bibinfo{volume}{79}, \bibinfo{number}{8} (\bibinfo{year}{1982}),
  \bibinfo{pages}{2554--2558}.
\newblock


\bibitem[Huang et~al\mbox{.}(2004)]%
        {cf_sparsity}
\bibfield{author}{\bibinfo{person}{Zan Huang}, \bibinfo{person}{Hsinchun Chen},
  {and} \bibinfo{person}{Daniel~Dajun Zeng}.} \bibinfo{year}{2004}\natexlab{}.
\newblock \showarticletitle{Applying associative retrieval techniques to
  alleviate the sparsity problem in collaborative filtering}.
\newblock \bibinfo{journal}{\emph{{ACM} Transactions on Information Systems}}
  \bibinfo{volume}{22}, \bibinfo{number}{1} (\bibinfo{year}{2004}),
  \bibinfo{pages}{116--142}.
\newblock


\bibitem[Ie et~al\mbox{.}(2019a)]%
        {slateQ:ijcai19}
\bibfield{author}{\bibinfo{person}{Eugene Ie}, \bibinfo{person}{Vihan Jain},
  \bibinfo{person}{Jing Wang}, \bibinfo{person}{Sanmit Narvekar},
  \bibinfo{person}{Ritesh Agarwal}, \bibinfo{person}{Rui Wu},
  \bibinfo{person}{Heng-Tze Cheng}, \bibinfo{person}{Tushar Chandra}, {and}
  \bibinfo{person}{Craig Boutilier}.} \bibinfo{year}{2019}\natexlab{a}.
\newblock \showarticletitle{{SlateQ}: A Tractable Decomposition for
  Reinforcement Learning with Recommendation Sets}. In
  \bibinfo{booktitle}{\emph{Proceedings of the Twenty-eighth International
  Joint Conference on Artificial Intelligence (IJCAI-19)}}.
  \bibinfo{address}{Macau}, \bibinfo{pages}{2592--2599}.
\newblock


\bibitem[Ie et~al\mbox{.}(2019b)]%
        {recsim:arxiv19}
\bibfield{author}{\bibinfo{person}{Eugene Ie}, \bibinfo{person}{Chih wei Hsu},
  \bibinfo{person}{Martin Mladenov}, \bibinfo{person}{Vihan Jain},
  \bibinfo{person}{Sanmit Narvekar}, \bibinfo{person}{Jing Wang},
  \bibinfo{person}{Rui Wu}, {and} \bibinfo{person}{Craig Boutilier}.}
  \bibinfo{year}{2019}\natexlab{b}.
\newblock \bibinfo{title}{RecSim: A Configurable Simulation Platform for
  Recommender Systems}.  (\bibinfo{year}{2019}).
\newblock
\newblock
\shownote{{\tt arXiv:1909.04847}}.


\bibitem[Keeney and Raiffa(1993)]%
        {keeney1993decisions}
\bibfield{author}{\bibinfo{person}{Ralph~L Keeney} {and}
  \bibinfo{person}{Howard Raiffa}.} \bibinfo{year}{1993}\natexlab{}.
\newblock \bibinfo{booktitle}{\emph{Decisions with multiple objectives:
  preferences and value trade-offs}}.
\newblock \bibinfo{publisher}{Cambridge university press}.
\newblock


\bibitem[Kohavi et~al\mbox{.}(2009)]%
        {kohavi_controlledAB:DMKD2009}
\bibfield{author}{\bibinfo{person}{Ron Kohavi}, \bibinfo{person}{Roger
  Longbotham}, \bibinfo{person}{Dan Sommerfield}, {and}
  \bibinfo{person}{Randal~M. Henne}.} \bibinfo{year}{2009}\natexlab{}.
\newblock \showarticletitle{Controlled Experiments on the Web: Survey and
  Practical Guide}.
\newblock \bibinfo{journal}{\emph{Data Mining and Knowledge Discovery}}
  \bibinfo{volume}{18} (\bibinfo{year}{2009}), \bibinfo{pages}{140--181}.
\newblock


\bibitem[Lai and Robbins(1985)]%
        {lai85asymptotically}
\bibfield{author}{\bibinfo{person}{T.~L. Lai} {and} \bibinfo{person}{Herbert
  Robbins}.} \bibinfo{year}{1985}\natexlab{}.
\newblock \showarticletitle{Asymptotically Efficient Adaptive Allocation
  Rules}.
\newblock \bibinfo{journal}{\emph{Advances in Applied Mathematics}}
  \bibinfo{volume}{6}, \bibinfo{number}{1} (\bibinfo{year}{1985}),
  \bibinfo{pages}{4--22}.
\newblock


\bibitem[Lam et~al\mbox{.}(2008)]%
        {Lam2008}
\bibfield{author}{\bibinfo{person}{Xuan~Nhat Lam}, \bibinfo{person}{Thuc Vu},
  \bibinfo{person}{Trong~Duc Le}, {and} \bibinfo{person}{Anh~Duc Duong}.}
  \bibinfo{year}{2008}\natexlab{}.
\newblock \showarticletitle{Addressing cold-start problem in recommendation
  systems}. In \bibinfo{booktitle}{\emph{Proceedings of the 2nd International
  Conference on Ubiquitous Information Management and Communication}}.
\newblock


\bibitem[Lattimore and Szepesvari(2020)]%
        {lattimore19bandit}
\bibfield{author}{\bibinfo{person}{Tor Lattimore} {and} \bibinfo{person}{Csaba
  Szepesvari}.} \bibinfo{year}{2020}\natexlab{}.
\newblock \bibinfo{booktitle}{\emph{Bandit Algorithms}}.
\newblock


\bibitem[Linden et~al\mbox{.}(2003)]%
        {Linden2003}
\bibfield{author}{\bibinfo{person}{Greg Linden}, \bibinfo{person}{Brent Smith},
  {and} \bibinfo{person}{Jeremy York}.} \bibinfo{year}{2003}\natexlab{}.
\newblock \showarticletitle{{Amazon}.com recommendations: Item-to-item
  collaborative filtering}.
\newblock \bibinfo{journal}{\emph{IEEE Distributed Systems Online}}
  \bibinfo{volume}{4} (\bibinfo{year}{2003}).
\newblock


\bibitem[Littman et~al\mbox{.}(2001)]%
        {littman_psr}
\bibfield{author}{\bibinfo{person}{Michael~L. Littman},
  \bibinfo{person}{Richard~S. Sutton}, {and} \bibinfo{person}{Satinder Singh}.}
  \bibinfo{year}{2001}\natexlab{}.
\newblock \showarticletitle{Predictive Representations of State}. In
  \bibinfo{booktitle}{\emph{Advances in Neural Information Processing Systems
  14 [Neural Information Processing Systems: Natural and Synthetic, {NIPS}
  2001, December 3-8, 2001, Vancouver, British Columbia, Canada]}}.
  \bibinfo{publisher}{{MIT} Press}, \bibinfo{pages}{1555--1561}.
\newblock


\bibitem[Liu et~al\mbox{.}(2016)]%
        {liu_etal:icdm16}
\bibfield{author}{\bibinfo{person}{Q. Liu}, \bibinfo{person}{S. Wu},
  \bibinfo{person}{D. Wang}, \bibinfo{person}{Z. Li}, {and} \bibinfo{person}{L.
  Wang}.} \bibinfo{year}{2016}\natexlab{}.
\newblock \showarticletitle{Context-Aware Sequential Recommendation}. In
  \bibinfo{booktitle}{\emph{Proceedings of the {IEEE} International Conference
  on Data Mining (ICDM-16)}}. \bibinfo{address}{Barcelona},
  \bibinfo{pages}{1053--1058}.
\newblock


\bibitem[Louviere et~al\mbox{.}(2000)]%
        {louviere-et-al:statedchoice2000}
\bibfield{author}{\bibinfo{person}{Jordan~J. Louviere},
  \bibinfo{person}{David~A. Hensher}, {and} \bibinfo{person}{Joffre~D. Swait}.}
  \bibinfo{year}{2000}\natexlab{}.
\newblock \bibinfo{booktitle}{\emph{Stated Choice Methods: Analysis and
  Application}}.
\newblock \bibinfo{publisher}{Cambridge University Press},
  \bibinfo{address}{Cambridge}.
\newblock


\bibitem[McNee et~al\mbox{.}(2003)]%
        {mcnee_onboarding:um03}
\bibfield{author}{\bibinfo{person}{Sean~M. McNee}, \bibinfo{person}{Shyong~K.
  Lam}, \bibinfo{person}{Joseph~A. Konstan}, {and} \bibinfo{person}{John
  Riedl}.} \bibinfo{year}{2003}\natexlab{}.
\newblock \showarticletitle{Interfaces for Eliciting New User Preferences in
  Recommender Systems}. In \bibinfo{booktitle}{\emph{Proceedings of the 9th
  International Conference on User Modeling (UM-03)}}.
  \bibinfo{address}{Johnstown, PA}, \bibinfo{pages}{178--187}.
\newblock


\bibitem[Meshi et~al\mbox{.}(2023)]%
        {meshi2023preference}
\bibfield{author}{\bibinfo{person}{Ofer Meshi}, \bibinfo{person}{Jon Feldman},
  \bibinfo{person}{Li Yang}, \bibinfo{person}{Ben Scheetz},
  \bibinfo{person}{Yanli Cai}, \bibinfo{person}{Mohammadhossein Bateni},
  \bibinfo{person}{Corbyn Salisbury}, \bibinfo{person}{Vikram Aggarwal}, {and}
  \bibinfo{person}{Craig Boutilier}.} \bibinfo{year}{2023}\natexlab{}.
\newblock \showarticletitle{Preference Elicitation for Music Recommendations}.
  In \bibinfo{booktitle}{\emph{ICML 2023 Workshop The Many Facets of
  Preference-Based Learning}}.
\newblock


\bibitem[Mladenov et~al\mbox{.}(2020)]%
        {mladenov_etal:icml20}
\bibfield{author}{\bibinfo{person}{Martin Mladenov}, \bibinfo{person}{Elliot
  Creager}, \bibinfo{person}{Kevin Swerksy}, \bibinfo{person}{Omer Ben-Porat},
  \bibinfo{person}{Richard~S. Zemel}, {and} \bibinfo{person}{Craig Boutilier}.}
  \bibinfo{year}{2020}\natexlab{}.
\newblock \showarticletitle{Optimizing Long-term Social Welfare in Recommender
  Systems: A Constrained Matching Approach}. In
  \bibinfo{booktitle}{\emph{Proceedings of the Thirty-seventh International
  Conference on Machine Learning (ICML-20)}}. \bibinfo{address}{Vienna},
  \bibinfo{pages}{6987--6998}.
\newblock


\bibitem[Mladenov et~al\mbox{.}(2021)]%
        {recsimNG:arxiv21}
\bibfield{author}{\bibinfo{person}{Martin Mladenov},
  \bibinfo{person}{{Chih-wei} Hsu}, \bibinfo{person}{Vihan Jain},
  \bibinfo{person}{Eugene Ie}, \bibinfo{person}{Chris Colby},
  \bibinfo{person}{Nic Mayoraz}, \bibinfo{person}{Hubert Pham},
  \bibinfo{person}{Dustin Tran}, \bibinfo{person}{Ivan Vendrov}, {and}
  \bibinfo{person}{Craig Boutilier}.} \bibinfo{year}{2021}\natexlab{}.
\newblock \bibinfo{title}{RecSim NG: Toward Principled Uncertainty Modeling for
  Recommender Ecosystems}.  (\bibinfo{year}{2021}).
\newblock
\newblock
\shownote{{\tt arXiv:2103.08057}}.


\bibitem[Pal et~al\mbox{.}(2020)]%
        {Pal2020}
\bibfield{author}{\bibinfo{person}{Aditya Pal}, \bibinfo{person}{Chantat
  Eksombatchai}, \bibinfo{person}{Yitong Zhou}, \bibinfo{person}{Bo Zhao},
  \bibinfo{person}{Charles Rosenberg}, {and} \bibinfo{person}{Jure Leskovec}.}
  \bibinfo{year}{2020}\natexlab{}.
\newblock \showarticletitle{{PinnerSage}: Multi-modal user embedding framework
  for recommendations at {Pinterest}}. In \bibinfo{booktitle}{\emph{Proceedings
  of the 26th ACM SIGKDD International Conference on Knowledge Discovery \&
  Data Mining}}.
\newblock


\bibitem[Pearl(2009)]%
        {pearl_book}
\bibfield{author}{\bibinfo{person}{Judea Pearl}.}
  \bibinfo{year}{2009}\natexlab{}.
\newblock \bibinfo{booktitle}{\emph{Causality: Models, Reasoning and Inference}
  (\bibinfo{edition}{2nd} ed.)}.
\newblock \bibinfo{publisher}{Cambridge University Press}.
\newblock


\bibitem[Peters et~al\mbox{.}(2017)]%
        {elements_of_causal_inference}
\bibfield{author}{\bibinfo{person}{Jonas Peters}, \bibinfo{person}{Dominik
  Janzing}, {and} \bibinfo{person}{Bernhard Schlkopf}.}
  \bibinfo{year}{2017}\natexlab{}.
\newblock \bibinfo{booktitle}{\emph{Elements of Causal Inference: Foundations
  and Learning Algorithms}}.
\newblock \bibinfo{publisher}{The MIT Press}.
\newblock


\bibitem[Quadrana et~al\mbox{.}(2017)]%
        {quadrana_etal:recsys17}
\bibfield{author}{\bibinfo{person}{Massimo Quadrana},
  \bibinfo{person}{Alexandros Karatzoglou}, \bibinfo{person}{Bal{\'{a}}zs
  Hidasi}, {and} \bibinfo{person}{Paolo Cremonesi}.}
  \bibinfo{year}{2017}\natexlab{}.
\newblock \showarticletitle{Personalizing Session-based Recommendations with
  Hierarchical Recurrent Neural Networks}. In
  \bibinfo{booktitle}{\emph{Proceedings of the Eleventh {ACM} Conference on
  Recommender Systems (RecSys-17)}}. \bibinfo{pages}{130--137}.
\newblock


\bibitem[Radensky et~al\mbox{.}(2023)]%
        {radensky_etal:facct23}
\bibfield{author}{\bibinfo{person}{Marissa Radensky},
  \bibinfo{person}{Julie~Anne S{\'{e}}guin}, \bibinfo{person}{Jang~Soo Lim},
  \bibinfo{person}{Kristen Olson}, {and} \bibinfo{person}{Robert Geiger}.}
  \bibinfo{year}{2023}\natexlab{}.
\newblock \showarticletitle{"I Think You Might Like This": Exploring Effects of
  Confidence Signal Patterns on Trust in and Reliance on Conversational
  Recommender Systems}. In \bibinfo{booktitle}{\emph{Proceedings of the 2023
  {ACM} Conference on Fairness, Accountability, and Transparency, FAccT 2023,
  Chicago, IL, USA, June 12-15, 2023}}. \bibinfo{pages}{792--804}.
\newblock


\bibitem[Rashid et~al\mbox{.}(2008)]%
        {AlMamunur2008}
\bibfield{author}{\bibinfo{person}{Al~Mamunur Rashid}, \bibinfo{person}{George
  Karypis}, {and} \bibinfo{person}{John Riedl}.}
  \bibinfo{year}{2008}\natexlab{}.
\newblock \showarticletitle{Learning preferences of new users in recommender
  systems: An information theoretic approach}.
\newblock \bibinfo{journal}{\emph{SIGKDD Explor. Newsl.}}  \bibinfo{volume}{10}
  (\bibinfo{year}{2008}), \bibinfo{pages}{90–100}.
\newblock


\bibitem[Rohde et~al\mbox{.}(2018)]%
        {rohde2018recogym}
\bibfield{author}{\bibinfo{person}{David Rohde}, \bibinfo{person}{Stephen
  Bonner}, \bibinfo{person}{Travis Dunlop}, \bibinfo{person}{Flavian Vasile},
  {and} \bibinfo{person}{Alexandros Karatzoglou}.}
  \bibinfo{year}{2018}\natexlab{}.
\newblock \bibinfo{title}{RecoGym: A Reinforcement Learning Environment for the
  problem of Product Recommendation in Online Advertising}.
  (\bibinfo{year}{2018}).
\newblock
\newblock
\shownote{{\tt arXiv:1808.00720 [cs.IR]}}.


\bibitem[Salakhutdinov and Mnih(2007)]%
        {salakhutdinov-mnih:nips07}
\bibfield{author}{\bibinfo{person}{Ruslan Salakhutdinov} {and}
  \bibinfo{person}{Andriy Mnih}.} \bibinfo{year}{2007}\natexlab{}.
\newblock \showarticletitle{Probabilistic Matrix Factorization}. In
  \bibinfo{booktitle}{\emph{Advances in Neural Information Processing Systems
  20 (NIPS-07)}}. \bibinfo{address}{Vancouver}, \bibinfo{pages}{1257--1264}.
\newblock


\bibitem[Salo and Hamalainen(2001)]%
        {salo2001preference}
\bibfield{author}{\bibinfo{person}{Ahti~A Salo} {and} \bibinfo{person}{Raimo~P
  Hamalainen}.} \bibinfo{year}{2001}\natexlab{}.
\newblock \showarticletitle{Preference ratios in multiattribute evaluation
  ({PRIME})-elicitation and decision procedures under incomplete information}.
\newblock \bibinfo{journal}{\emph{IEEE Transactions on Systems, Man, and
  Cybernetics-Part A: Systems and Humans}}  \bibinfo{volume}{31}
  (\bibinfo{year}{2001}), \bibinfo{pages}{533--545}.
\newblock


\bibitem[Shani and Gunawardana(2011)]%
        {shani:RSHandbook2011}
\bibfield{author}{\bibinfo{person}{Guy Shani} {and} \bibinfo{person}{Asela
  Gunawardana}.} \bibinfo{year}{2011}\natexlab{}.
\newblock \showarticletitle{Evaluating Recommendation Systems}.
\newblock In \bibinfo{booktitle}{\emph{Recommender Systems Handbook}},
  \bibfield{editor}{\bibinfo{person}{F.~Ricci}, \bibinfo{person}{L.~Rokach},
  \bibinfo{person}{B.~Shapira}, {and} \bibinfo{person}{P.~Kantor}} (Eds.).
  \bibinfo{publisher}{Springer}, \bibinfo{address}{Boston},
  \bibinfo{pages}{257--297}.
\newblock


\bibitem[van~den Oord et~al\mbox{.}(2013)]%
        {oord_etal:nips13}
\bibfield{author}{\bibinfo{person}{A{\"{a}}ron van~den Oord},
  \bibinfo{person}{Sander Dieleman}, {and} \bibinfo{person}{Benjamin
  Schrauwen}.} \bibinfo{year}{2013}\natexlab{}.
\newblock \showarticletitle{Deep content-based music recommendation}. In
  \bibinfo{booktitle}{\emph{Advances in Neural Information Processing Systems
  26: 27th Annual Conference on Neural Information Processing Systems 2013.
  Proceedings of a meeting held December 5-8, 2013, Lake Tahoe, Nevada, United
  States}}. \bibinfo{pages}{2643--2651}.
\newblock


\bibitem[Vaswani et~al\mbox{.}(2017a)]%
        {vaswani2017attention}
\bibfield{author}{\bibinfo{person}{Ashish Vaswani}, \bibinfo{person}{Noam
  Shazeer}, \bibinfo{person}{Niki Parmar}, \bibinfo{person}{Jakob Uszkoreit},
  \bibinfo{person}{Llion Jones}, \bibinfo{person}{Aidan~N Gomez},
  \bibinfo{person}{{\L}ukasz Kaiser}, {and} \bibinfo{person}{Illia
  Polosukhin}.} \bibinfo{year}{2017}\natexlab{a}.
\newblock \showarticletitle{Attention is all you need}.
\newblock \bibinfo{journal}{\emph{Advances in neural information processing
  systems}}  \bibinfo{volume}{30} (\bibinfo{year}{2017}).
\newblock


\bibitem[Vaswani et~al\mbox{.}(2017b)]%
        {vaswani_2017}
\bibfield{author}{\bibinfo{person}{Ashish Vaswani}, \bibinfo{person}{Noam
  Shazeer}, \bibinfo{person}{Niki Parmar}, \bibinfo{person}{Jakob Uszkoreit},
  \bibinfo{person}{Llion Jones}, \bibinfo{person}{Aidan~N Gomez},
  \bibinfo{person}{\L~ukasz Kaiser}, {and} \bibinfo{person}{Illia Polosukhin}.}
  \bibinfo{year}{2017}\natexlab{b}.
\newblock \showarticletitle{Attention is All you Need}. In
  \bibinfo{booktitle}{\emph{Advances in Neural Information Processing
  Systems}}, \bibfield{editor}{\bibinfo{person}{I.~Guyon},
  \bibinfo{person}{U.~Von Luxburg}, \bibinfo{person}{S.~Bengio},
  \bibinfo{person}{H.~Wallach}, \bibinfo{person}{R.~Fergus},
  \bibinfo{person}{S.~Vishwanathan}, {and} \bibinfo{person}{R.~Garnett}}
  (Eds.), Vol.~\bibinfo{volume}{30}. \bibinfo{publisher}{Curran Associates,
  Inc.}
\newblock
\urldef\tempurl%
\url{https://proceedings.neurips.cc/paper_files/paper/2017/file/3f5ee243547dee91fbd053c1c4a845aa-Paper.pdf}
\showURL{%
\tempurl}


\bibitem[Viappiani and Boutilier(2010)]%
        {viappiani:nips2010}
\bibfield{author}{\bibinfo{person}{Paolo Viappiani} {and}
  \bibinfo{person}{Craig Boutilier}.} \bibinfo{year}{2010}\natexlab{}.
\newblock \showarticletitle{Optimal {Bayesian} Recommendation Sets and
  Myopically Optimal Choice Query Sets}. In \bibinfo{booktitle}{\emph{Advances
  in Neural Information Processing Systems 23 (NIPS)}}.
  \bibinfo{address}{Vancouver}, \bibinfo{pages}{2352--2360}.
\newblock


\bibitem[Wang et~al\mbox{.}(2023)]%
        {wang_etal:recsys23}
\bibfield{author}{\bibinfo{person}{Yueqi Wang}, \bibinfo{person}{Yoni Halpern},
  \bibinfo{person}{Shuo Chang}, \bibinfo{person}{Jingchen Feng},
  \bibinfo{person}{Elaine~Ya Le}, \bibinfo{person}{Longfei Li},
  \bibinfo{person}{Xujian Liang}, \bibinfo{person}{Min{-}Cheng Huang},
  \bibinfo{person}{Shane Li}, \bibinfo{person}{Alex Beutel},
  \bibinfo{person}{Yaping Zhang}, {and} \bibinfo{person}{Shuchao Bi}.}
  \bibinfo{year}{2023}\natexlab{}.
\newblock \showarticletitle{Learning from Negative User Feedback and Measuring
  Responsiveness for Sequential Recommenders}. In
  \bibinfo{booktitle}{\emph{Proceedings of the 17th {ACM} Conference on
  Recommender Systems (RecSys-23)}}. \bibinfo{address}{Singapore},
  \bibinfo{pages}{1049--1053}.
\newblock


\bibitem[Yang et~al\mbox{.}(2020)]%
        {yangEtAl:www20}
\bibfield{author}{\bibinfo{person}{Ji Yang}, \bibinfo{person}{Xinyang Yi},
  \bibinfo{person}{Derek Zhiyuan~Cheng}, \bibinfo{person}{Lichan Hong},
  \bibinfo{person}{Yang Li}, \bibinfo{person}{Simon Xiaoming~Wang},
  \bibinfo{person}{Taibai Xu}, {and} \bibinfo{person}{Ed~H. Chi}.}
  \bibinfo{year}{2020}\natexlab{}.
\newblock \showarticletitle{Mixed Negative Sampling for Learning Two-tower
  Neural Networks in Recommendations}. In \bibinfo{booktitle}{\emph{Proceedings
  of the Web Conference (WWW-20)}}. \bibinfo{address}{Taipei},
  \bibinfo{pages}{441--447}.
\newblock


\bibitem[Yao et~al\mbox{.}(2020)]%
        {yao_halpern_etal:fates20}
\bibfield{author}{\bibinfo{person}{Siriu Yao}, \bibinfo{person}{Yoni Halpern},
  \bibinfo{person}{Nithum Thain}, \bibinfo{person}{Xuezhi Wang},
  \bibinfo{person}{Kang Lee}, \bibinfo{person}{Flavien Prost},
  \bibinfo{person}{Ed~H. Chi}, \bibinfo{person}{Jilin Chen}, {and}
  \bibinfo{person}{Alex Beutel}.} \bibinfo{year}{2020}\natexlab{}.
\newblock \showarticletitle{Measuring Recommender System Effects with Simulated
  Users}. In \bibinfo{booktitle}{\emph{2nd Workshop on Fairness,
  Accountability, Transparency, Ethics and Society on the Web (FATES-20)}}.
  \bibinfo{address}{Taipei}.
\newblock


\bibitem[Yuan et~al\mbox{.}(2019)]%
        {yuan_etal:wsdm19}
\bibfield{author}{\bibinfo{person}{Fajie Yuan}, \bibinfo{person}{Alexandros
  Karatzoglou}, \bibinfo{person}{Ioannis Arapakis}, \bibinfo{person}{Joemon~M.
  Jose}, {and} \bibinfo{person}{Xiangnan He}.} \bibinfo{year}{2019}\natexlab{}.
\newblock \showarticletitle{A Simple Convolutional Generative Network for Next
  Item Recommendation}. In \bibinfo{booktitle}{\emph{Proceedings of the Twelfth
  {ACM} International Conference on Web Search and Data Mining (WSDM-19),
  {WSDM} 2019, Melbourne, VIC, Australia, February 11-15, 2019}}.
  \bibinfo{address}{Melbourne}, \bibinfo{pages}{582--590}.
\newblock


\bibitem[Zhao et~al\mbox{.}(2018)]%
        {zhao_slateRL:recsys18}
\bibfield{author}{\bibinfo{person}{Xiangyu Zhao}, \bibinfo{person}{Long Xia},
  \bibinfo{person}{Liang Zhang}, \bibinfo{person}{Zhuoye Ding},
  \bibinfo{person}{Dawei Yin}, {and} \bibinfo{person}{Jiliang Tang}.}
  \bibinfo{year}{2018}\natexlab{}.
\newblock \showarticletitle{Deep Reinforcement Learning for Page-wise
  Recommendations}. In \bibinfo{booktitle}{\emph{Proceedings of the 12th {ACM}
  Conference on Recommender Systems (RecSys-18)}}.
  \bibinfo{address}{Vancouver}, \bibinfo{pages}{95--103}.
\newblock


\end{thebibliography}
\newpage
\appendix
\section{Assumptions for Session Linearization}
\label{sec:app}

We should note that the data format for training \emph{user session generator} in Section \ref{sec:session_gen} makes certain simplifying assumptions, since the onboarding UI does not in fact ask about a single artist at a time, but rather displays multiple artists/queries at the same time (with a slate size that varies with device). See \cref{fig:TB_screen}. We make a simplifying assumption about user interactions to allow training of more tractable sequence models of user behavior, specifically, adopting something akin to a \emph{cascade model} \cite{cascade_model}:
\begin{itemize}
    \item The user examines artists in a linear fashion when multiple artists are displayed, and scrolls to the next screen when all displayed artists are examined (assuming non-termination); the user never scrolls back to reexamine artists.
    \item Once examined, the user decides whether to select or skip that artist. The user never reverts their decisions (e.g., by deselecting an artist). We also assume the user must examine all artists displayed even if they left the OP immediately.
    \item New artists can be displayed ``on-the-fly'' (insertion) given any previous selection. We assume the user examines these new artists immediately after the selection that generated them, even if other artists were displayed with the selected artist.
\end{itemize}
The linearization assumption is less realistic with the YouTube Music desktop app (which shows more artists per screen) than the mobile app; so we confine our attention in this work to mobile users.

\section{Additional Results}
\label{sec:additional_results}
Table~\ref{tab:LEvsLaunch} shows the policy ordering induced by  the three small ``true LEs'', the ``Launch'', and the simulation. For completeness, Table~\ref{tab:LEvsLaunchImpression} shows similar results for artist impressions. Notice that none of the three ``true LEs'' order the policies in the same way. Indeed, none of the ``true LEs,'' nor the simulation test, predict the policy ordering of the ``Launch.'' The small LEs are slightly better aligned with the Launch than Simulation for $\lambda$ values of $0.001$ and $0.05$, in part, due to the variance in simulation results across all device and geo types. We note, however, the the LE and Launch are more correlated in our design than they would be in practice, since each LE is in fact a sample of 1/3 of the Launch traffic.

\begin{table}
\centering
\begin{small}
\begin{tabular}{|c||c|c|c|c|}
\hline
           & Imps. $\lambda=0.001$ & Imps. $\lambda=0.05$ & Imps. $\lambda=0.2$ & Ordering \\
\hline \hline
LE 1 & $0.18 \pm 1.29\% $ & $-0.44 \pm 1.08\% $ & $0.53 \pm 1.52\% $ & $0.2 \!\!>\!\! 0.001 \!\!>\!\! 0.05$ \\
LE 2 & $-0.62 \pm 1.53\% $ & $0.19 \pm 1.86\% $ & $-0.57 \pm 1.62\% $ & $0.05 \!\!>\!\! 0.2 \!\!>\!\! 0.001$ \\ 
LE 3 & $-0.55 \pm 1.55\% $ & $-0.49 \pm 1.19\% $ & $0.72 \pm 1.67\% $ & $0.2 \!\!>\!\! 0.05 \!\!>\!\! 0.001$ \\
Launch & $-0.29 \pm 0.72\% $ & $-0.29 \pm 0.89\% $ & $0.30 \pm 1.13\% $ & $0.05 \!\!>\!\! 0.001 \!\!>\!\! 0.2$ \\
Simul. & $0.75 \pm 0.83\% $ & $0.47 \pm 0.70\% $ & $0.50 \pm 0.83\% $ & $0.001 \!\!>\!\! 0.2 \!\!>\!\! 0.05$ \\ \hline
\end{tabular}
\end{small}
\caption{Impression Change (95\% CIs) for 3 Policies, and Induced Policy Ordering: LEs vs.\ ``Launch'' vs. Simulation.}
\label{tab:LEvsLaunchImpression}
\end{table}

\end{document}